\documentclass[preprint,12pt]{elsarticle}
\usepackage{graphicx}
\usepackage{amsmath}
\usepackage{amssymb}
\usepackage{amsthm}
\usepackage{caption}
\usepackage{subcaption}
\usepackage{xcolor}
\usepackage{bm}

\usepackage{ulem}
\usepackage{placeins}
\usepackage{multicol}
\usepackage{multirow}
\usepackage{adjustbox}
\usepackage[ruled, linesnumbered]{algorithm2e}
\SetKwInput{KwInput}{Input}                
\SetKwInput{KwOutput}{Output}              
\usepackage{setspace} 
\begin{document}
\journal{Aerospace Science and Technology}

\begin{frontmatter}

\title{Intermittency of a transitional airfoil flow \\ with laminar separation bubble \\ solved by the lattice-Boltzmann method}

\author[inst1]{Bernardo Luiz Ribeiro}
\author[inst1]{Cayan Dantas}
\author[inst1]{William Wolf}

\affiliation[inst1]{organization={University of Campinas},
            city={Campinas},
            postcode={13086-860},
            country={Brazil}}

\begin{abstract}
The flow over a NACA0012 airfoil at a moderate Reynolds number $Re=5 \times 10^4$ and angle of attack of $\alpha = 3^{\circ}$ is investigated using the lattice-Boltzmann method (LBM). The LBM solutions are computed in direct numerical simulation (DNS) mode, i.e., without a wall model. A validation is performed against a Navier-Stokes wall-resolved large eddy simulation, and good agreement is achieved between the different approaches, showing that the LBM can provide accurate solutions of boundary layers under transitional regime, but with a significant computational cost reduction. A laminar separation bubble (LSB) forms over the suction side of the airfoil, leading to intermittent vortex shedding that impacts transition to turbulence and the generation of strong spanwise-coherent vortices. Different shedding patterns are observed including the advection of single vortical structures and pairing of two vortices, which may or may not break into finer turbulent scales. Such flow features are characterized by 2D and 3D events that directly impact the sound generation by the trailing edge. Frequency and amplitude modulations from the LSB lead to a noise spectrum with a main tone plus equidistant secondary tones, and a time-frequency analysis shows that the main tones may switch frequencies due to intermittency. This research advances in the comprehension of the LSB behavior in transitional airfoil flows, impacting the performance and noise generation of blades and propellers.
\end{abstract}


\end{frontmatter}

\section{Introduction}

The investigation of airfoil flows at low and moderate Reynolds numbers ($10^4 < Re < 10^6$) finds application in the design of quieter micro-air vehicles (MAVs) and electric vertical take-off and landing (eVTOL) aircraft. 
For instance, recent experimental and numerical studies have shown that propellers operating in the transitional flow regime develop laminar separation bubbles (LSBs) which impact the unsteady flow dynamics and the subsequent trailing-edge (TE) noise generation \cite{casalino5, aescte2, aescte5, aescte1, aescte4}. 

Airfoil noise has been extensively investigated for a range of Reynolds numbers, including laminar, transitional, and turbulent regimes, on different airfoil profiles at various angles of attack \cite{paterson, Arbey1983, Brooks1989, moreau5, wolf2012, probsting2015, ricciardi2022PRF, Sano_2023}. Depending on the Reynolds number, different mechanisms drive the airfoil self-noise generation. At values of $Re < 10^5$, the flow dynamics over the airfoil suction side dominates over the pressure side, whereas at high Reynolds numbers the opposite becomes true \cite{probsting2015, ricciardi2022PRF}. 

A major feature of low and moderate Reynolds number flows consists in the formation of an LSB on the airfoil suction side. This bubble is shown to modulate the shedding frequency and amplitude of coherent structures advected toward the airfoil TE and such modulations lead to pressure spectra with multiple equidistant tones \cite{paterson, Arbey1983, Desquesnes2007, Ricciardi2019_tones}. Pröbsting and Yarusevych \cite{probsting_yaru2015} found that a NACA0012 airfoil at Reynolds numbers $0.65 \times 10^5 \leq Re \leq 4.5 \times 10^5$ and angle of attack $\alpha = 2^{\circ}$ presents an intermittent laminar-turbulent transition that affects the convection of coherent structures from the LSB over the suction side. In these cases, the TE scattering of spanwise-coherent structures is a major mechanism for airfoil tonal noise generation \cite{expfluids, SanoPRF2019, moreau1}.

Recently, Ricciardi et al. \cite{ricciardiJFM} revealed that the vortex merging and bursting over the airfoil suction side depends on the constructive phase interference between the unstable frequencies observed in the eigenspectrum computed from a global stability analysis. When the frequencies are in phase, it is likely that a successful event of vortex pairing occurs; otherwise, vortex bursting might happen leading to uncorrelated turbulent structures. These authors also showed that the noise spectrum depicts a main tone with multiple equidistant secondary tones, in agreement with previous numerical and experimental studies \cite{paterson, Arbey1983, Desquesnes2007, expfluids, probsting2015}. The acoustic disturbances emitted at the TE also close an acoustic feedback loop  \cite{Desquesnes2007, jones2008, FosasdePando2014} when reaching the most receptive region near the leading edge, as demonstrated by Ref. \cite{ricciardiJFM}. The pairing dynamics of the shed vortices has also been explained by an empirical mode decomposition \cite{lucas2024}. The previous authors found that the successful pairing depends on changes in the convective speeds of two coherent vortices shed by the LSB with similar sizes, where the upstream vortex accelerates, while the downstream one decelerates. Dynamic systems theory was employed by Sano et al. \cite{Sano_2023} to demonstrate that the secondary tones emerge at low Reynolds numbers as a switch from quasi-periodic to chaotic dynamics near the trailing edge.

Most numerical investigations of the flow dynamics and aeroacoustics of airfoils at moderate Reynolds numbers have been carried out solving the 2D \cite{Desquesnes2007, jones2008, FosasdePando2014, ricciardi2019} and 3D \cite{jones2008, moreau5, ricciardiJFM, ricciardi2022PRF} compressible Navier-Stokes (NS) equations. However, since the 2000s, the lattice-Boltzmann method (LBM) became an alternative for unsteady flow simulations with separation. More recently, this approach has also been applied for the study of attached flows over airfoils at moderate Reynolds numbers and transitional boundary layers with LSBs. Applications including spanwise-periodic wings and realistic propeller configurations have been presented in Refs. \cite{moreau5, aescte1, aescte2, aescte4}.
Due to its simpler partial differential equation system compared to the traditional NS solvers, the LBM offers advantages in terms of simulation time and scalability. Although Sanjose et al. \cite{moreau5} performed LBM calculations without a wall-model, in a direct numerical simulation (DNS) mode, most studies of transitional airfoil flows available in the literature employed very large-eddy simulations (VLES), where a wall-model is applied.


The objective of this work is to perform an assessment of the LBM in resolving a transitional airfoil flow with spanwise-periodic boundary conditions and no external disturbances. 
The simulations are run in DNS mode, i.e., without subgrid scale or wall models which are available in the present solver.  
Results are compared to a wall-resolved LES of the compressible NS equations from Ref. \cite{ricciardiJFM} in terms of mean and root-mean-square (RMS) flow statistics, besides spectral analyses. A mesh refinement study is also performed for the LBM simulations and, after validation of results, an investigation of the intermittency effects introduced by the LSB and its vortex shedding is presented. Differently from Ref. \cite{ricciardiJFM}, the focus of this study is on the spatiotemporal intermittency characterization of the LSB and its role on the airfoil near-wall dynamics. The understanding of the LSB behavior and its subsequent vortex shedding is essential to identify trailing-edge noise source mechanisms in transitional airfoils, as it provides insights into mitigation strategies, useful for the design of quieter urban air mobility. 

 \section{Methodology}
\subsection{Numerical simulations}

Numerical simulations are conducted using the SIMULIA PowerFLOW 6-2022 LBM-based code. The LBM resolves fluid flows at the mesoscopic scale in a statistical sense, having origin in the lattice gas models \cite{timmkruger}. While these models track the behavior of particles at a microscopic scale, the LBM instead tracks the advection and collision of fluid particles using discrete distribution functions $f_i(\bm{x}, t)$, often called particle populations \cite{aescte5}. The term $f_i$ represents the density of particles travelling with velocity $\bm{c}=(c_x, c_y, c_z)$ from the position $\bm{x}=(x, y, z)$ at a time $t$ in the direction $i$. The velocity is chosen so that particles travel one cell per timestep, effectively making the Courant–Friedrichs–Lewy (CFL) number for $f_i$ equal to one. The LBM equation is written as 

\begin{equation} \label{lbm}
    f_i(\bm{x}+\bm{c}_i \Delta t, t+\Delta t) - f_i(\bm{x}, t) = \Omega_i(\bm{x}, t) \mbox{ ,}
\end{equation}

\noindent where $\Omega_i$ represents the collision operator in the $i$-th direction and $\Delta t$ is the timestep. The left-hand side of Eq. (\ref{lbm}) represents particles moving with velocity $\bm{c}$ in the $i$-th direction to a neighbouring point $\bm{x}+\bm{c}_i \Delta t$ at the next timestep $t+\Delta t$. On the right-hand side, particle collisions are modeled by redistributing them among $f_i$ at each site. In PowerFLOW, the collision term is modeled using the approximation described by Bhatnagar, Gross and Krook (BGK) \cite{bgk} as
\begin{equation} \label{bgk}
    \Omega_i(\bm{x}, t)=-\frac{\Delta t}{\tau} [f_i(\bm{x}, t)-f_i^{eq}(\bm{x}, t)] \mbox{ ,}
\end{equation}

\noindent where $\tau$ is the relaxation time, which is related to the fluid kinematic viscosity $\nu$ through the relation $\nu=c_s^{2}(\tau-\Delta t/2)$. Here, $c_s$ is the speed of sound, and the term $f_i^{eq}$ is the equilibrium Maxwell-Boltzmann distribution function \cite{chen}, which is approximated by a second-order expansion as 
\begin{equation} \label{edb}
    f_i^{eq}= w_i \rho \left(1 + \frac{\bm{c}_i \cdot \bm{u}} {c_s^{2}} + \frac{(\bm{c}_i \cdot \bm{u})^{2}}{2c_s^{4}} - \frac{\bm{u} \cdot \bm{u}}{2c_s^{2}} \right) \mbox{ ,}
\end{equation}

\noindent where $w_i$ represents weighting coefficients which depend on the direction being calculated, while $\rho$ is the fluid density and $\bm{u}$ is the fluid macroscopic velocity. Together with a corresponding set of weights $w_i$, $\bm{c}_i$ forms velocity sets usually denoted by D\textit{d}Q\textit{q}, where \textit{d} is the number of spatial dimensions of the velocity set, and \textit{q} represents the number of discrete velocity directions. In PowerFLOW, the D3Q19 velocity set is used for three-dimensional flow simulations \cite{aescte2}. 

The phenomenon of transition to turbulence requires a level of spatial and temporal resolution that is compatible with DNS or wall-resolved LES. The turbulence modeling capability of PowerFLOW which enables a numerical procedure called very large-eddy simulation (VLES)  \cite{aescte1}, has not been shown to capture laminar-turbulence transition for low-Reynolds-number flows at undisturbed conditions, at least for previous PowerFLOW versions \cite{aescte4}. Therefore, in the present work we perform simulations in the DNS mode, i.e., we do not employ subgrid or wall models available in the solver. 

The link between Eq. (\ref{lbm}) and the Navier-Stokes equations can be determined using the Chapman-Enskog analysis \cite{chapman}. Once $f_i$ is known, macroscopic flow variables, such as the density $\rho$ and $\bm{u}$ are obtained by taking the zeroth- and first-order moments of $f_i$, respectively: 
\begin{equation} \label{sums1}
    \rho (\bm{x}, t)= \sum_{i=0}^{q-1} f_i(\bm{x}, t) \mbox{ ,}
\end{equation}

\noindent and

\begin{equation} \label{sums2}
    \rho (\bm{x}, t) \bm{u}(\bm{x}, t)=\sum_{i=0}^{q-1} \bm{c}_i f_i(\bm{x}, t) \mbox{ .}
\end{equation}
The LBM is solved on a lattice composed of cubical elements called voxels. The simulation domain can be subdivided into several regions where different voxel resolutions are applied \cite{moreau5}, such that the resolution between two adjacent regions varies by a factor of 2. Solid boundaries are discretized using computational surface elements (surfels) that are generated at locations where the voxels intersect solid surfaces (facets). The process of generating voxels and surfels is fully automated in the solver. The boundary condition at a solid wall is computed by applying appropriate particle interactions in the collision term of the LBM. Results are transient and time accurate, where the time advancement is performed by an explicit scheme which allows for efficient and highly-scalable simulations. 

In the present work, the LBM results are validated against the NS solutions from Ricciardi et al. \cite{ricciardiJFM}. In this previous reference, a wall-resolved LES was performed to solve the compressible NS equations in general curvilinear coordinates. 
The spatial discretization of the governing equations employs a sixth-order accurate compact scheme for derivatives and interpolations on a staggered grid. The time integration is carried out by a hybrid implicit–explicit method, where an implicit second-order scheme is applied in the near-wall region, while the outer region employs a third-order Runge-Kutta scheme. Away from the wall, a sixth-order compact filter is applied. No-slip adiabatic wall boundary conditions are enforced along the airfoil surface and characteristic plus sponge boundary conditions are applied in the far-field locations to minimize wave reflections. Periodic boundary conditions are used in the spanwise direction. Further details of the LES procedure can be found in Ref. \cite{ricciardiJFM}.


\subsection{Flow and mesh configurations} \label{s:mesh}

\begin{figure}[h!] 
    \centering
    \includegraphics[width = 0.7\textwidth]{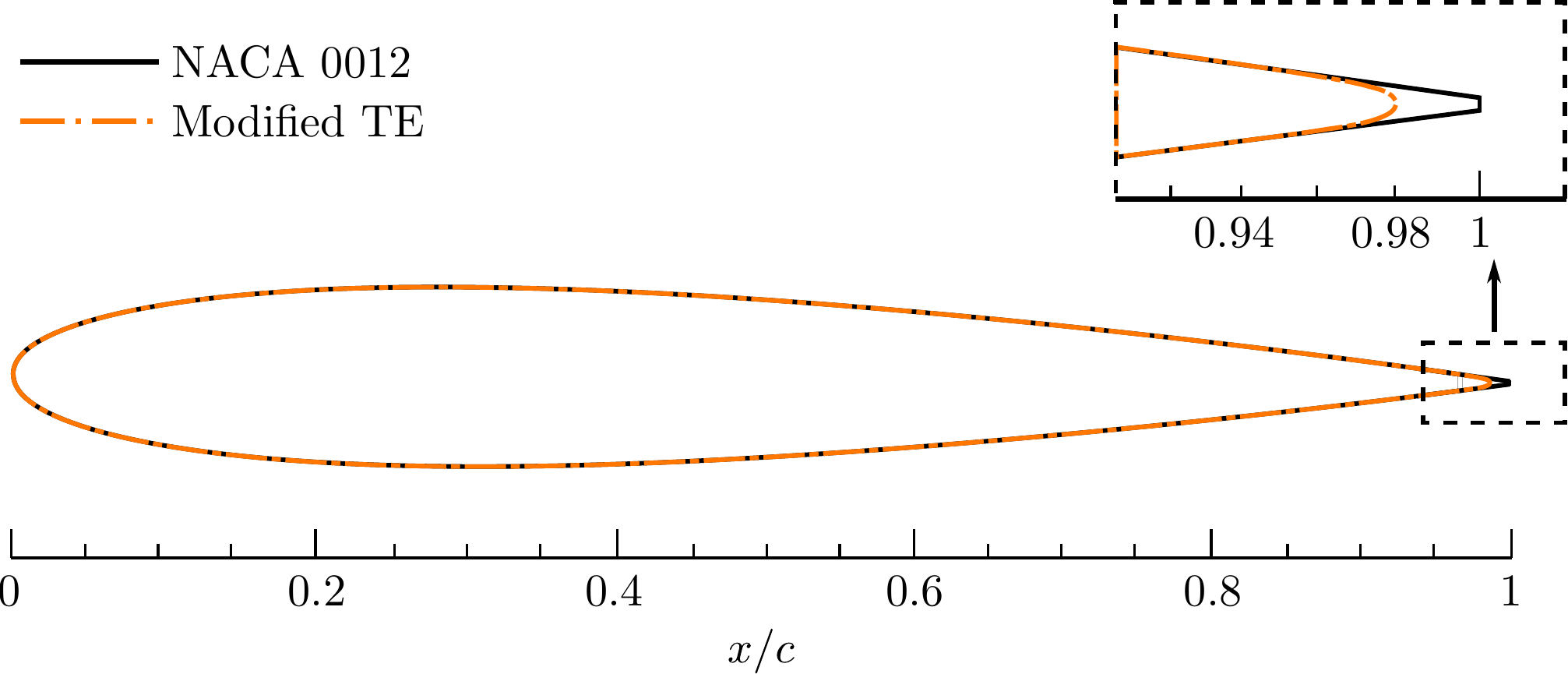}
    \caption{Modified NACA0012 TE profile from Ricciardi et al. \cite{ricciardiJFM}.}
    \label{naca}
\end{figure}

The flow over a NACA0012 airfoil is investigated for a Reynolds number $Re = 5 \times 10^4$, freestream Mach number $M_\infty = 0.3$, and angle of attack $\alpha = 3^{\circ}$. 
In order to allow a proper comparison with Ref. \cite{ricciardiJFM}, the trailing edge is modified as shown in Fig. \ref{naca}, and a periodic spanwise boundary condition is applied in the LBM.
The airfoil has a chord length $c$ and results are presented in nondimensional form using this reference length scale. 
Different mesh refinements are performed and according to the number of voxels over the airfoil, distinct spanwise lengths are achieved, ranging from approximately $30\%$ of the airfoil chord in the finest mesh to $50\%$ in the coarsest. The baseline mesh has the same spanwise length of the NS solution, i.e., $40\%$ of the chord.
Outlet boundary conditions are applied on the right, top and bottom planes of the computational domain, while an inlet is defined in the left plane which is the surface closest to the leading edge. The farfield boundary conditions are located $30$ chords away from the center of the profile on each direction. A no-slip boundary condition is specified at the airfoil surface and the simulation is conducted without any wall model to assess the capability of the LBM to capture the laminar to turbulent transition of the boundary layer. 

\FloatBarrier

\begin{figure}[h!]
\centering
\begin{subfigure}{0.49\textwidth}
    \includegraphics[trim={0cm 0cm 0cm 0cm},clip, width=1\linewidth]{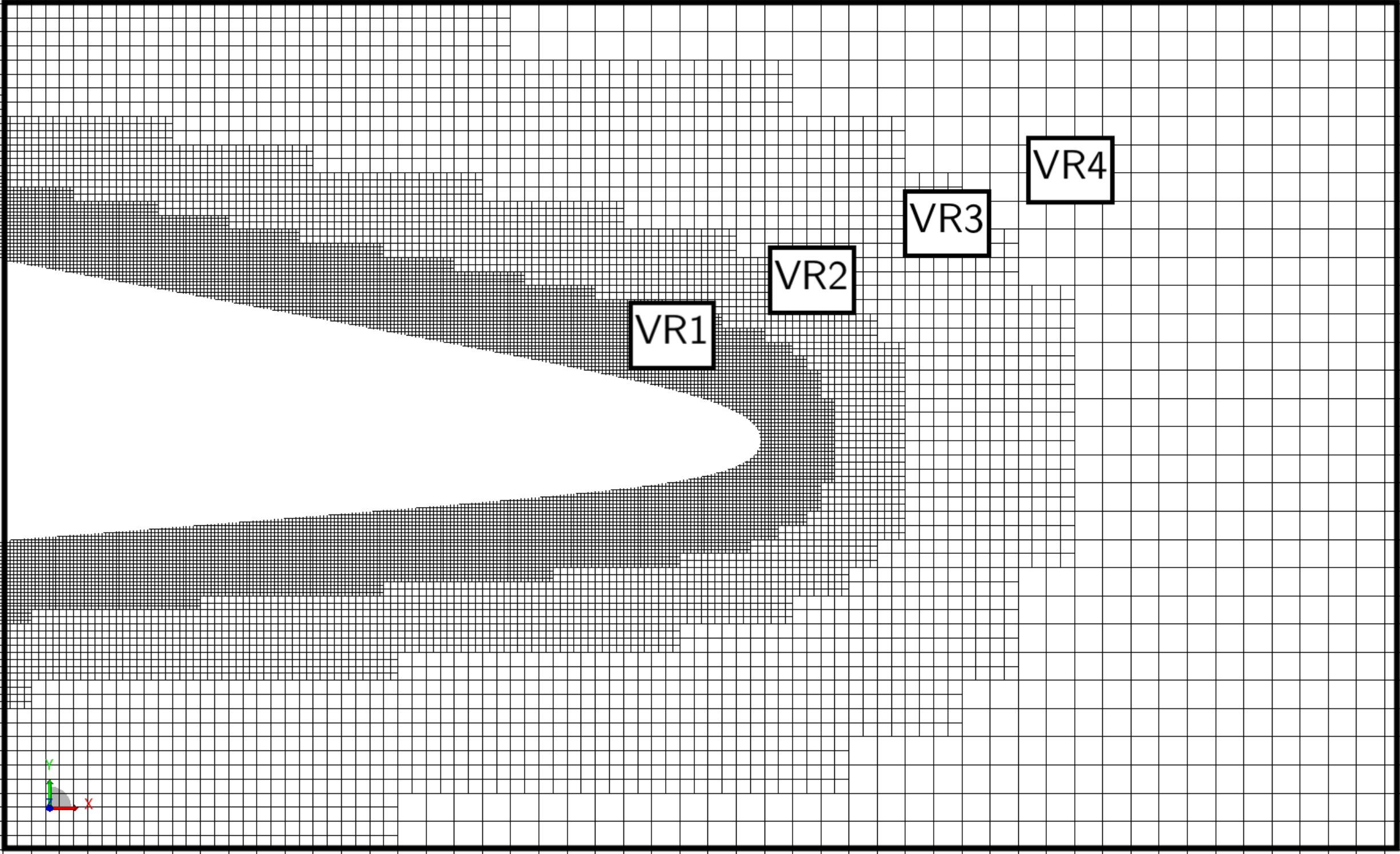}
    \caption{Coarse mesh.}
    \label{f:coarse1}
\end{subfigure}
\begin{subfigure}{0.476\textwidth}
    \includegraphics[trim={0cm 0cm 0cm 0cm},clip, width=1\linewidth]{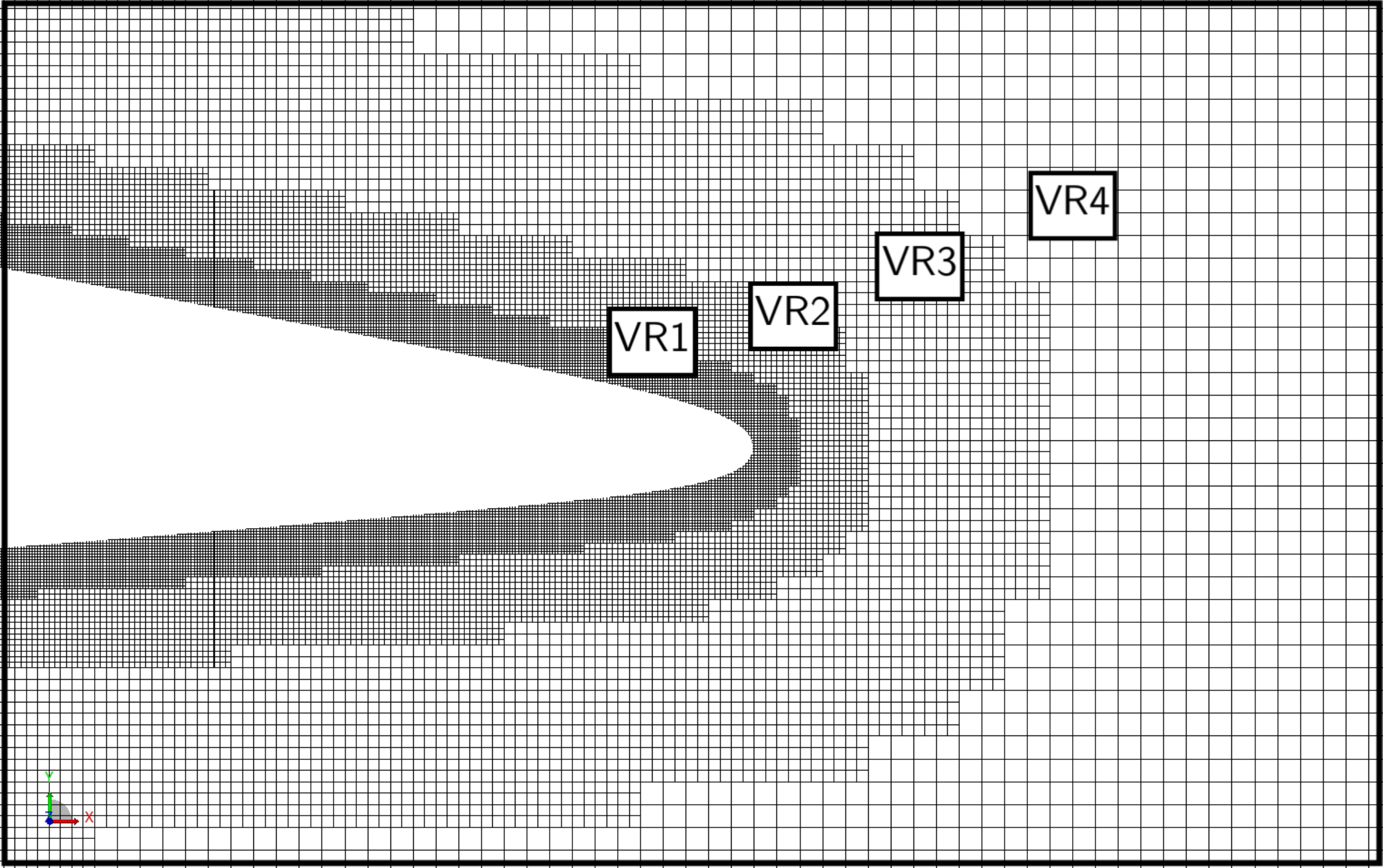}
    \caption{Baseline mesh.}
    \label{f:baseline1}
\end{subfigure}

\begin{subfigure}{0.48\textwidth}
    \includegraphics[trim={0cm 0cm 0cm 0cm},clip, width=1\linewidth]{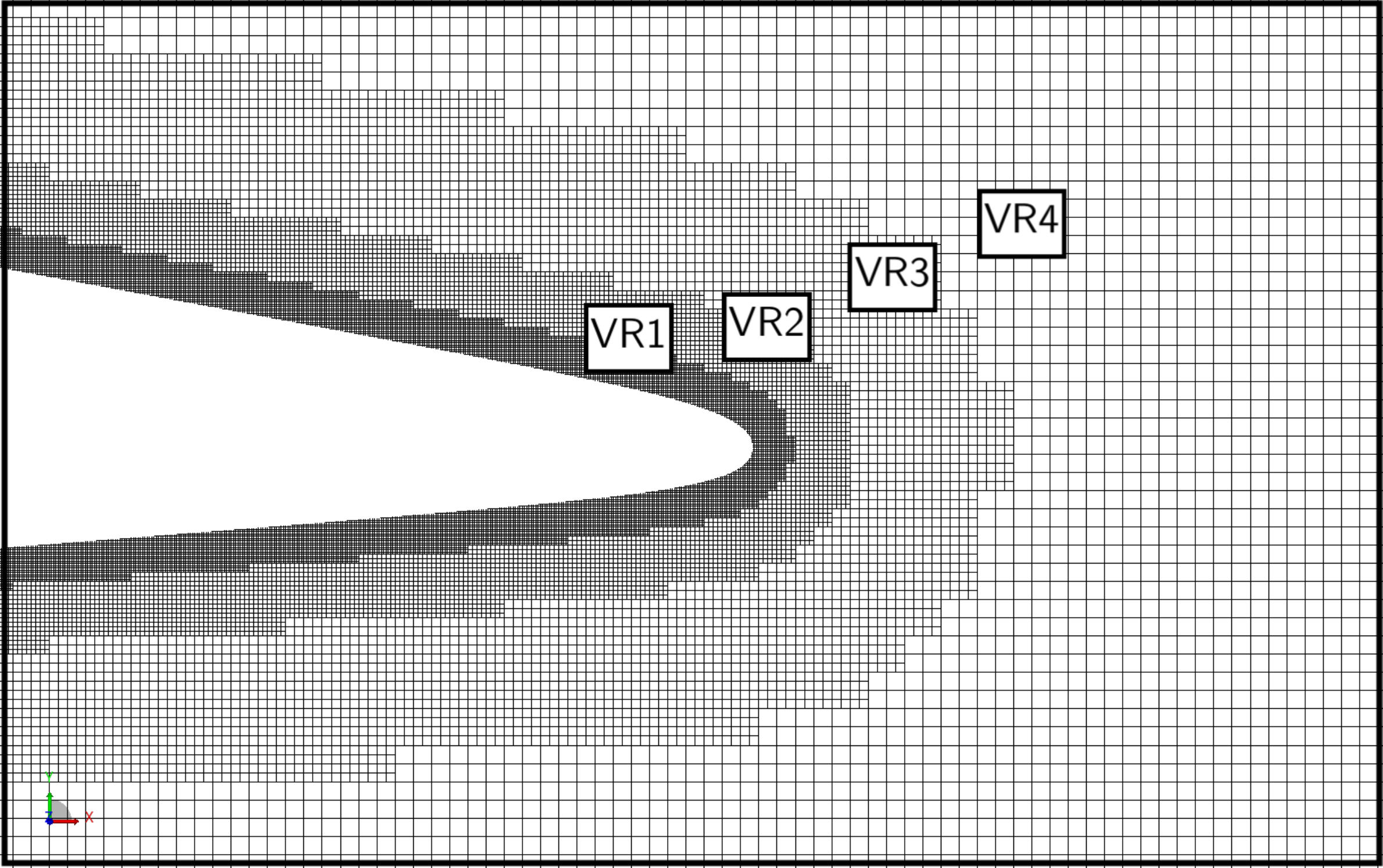}
    \caption{Fine mesh.}
    \label{f:fine1}
\end{subfigure}
\begin{subfigure}{0.485\textwidth}
    \includegraphics[trim={0cm 0cm 0cm 0cm},clip, width=1\linewidth]{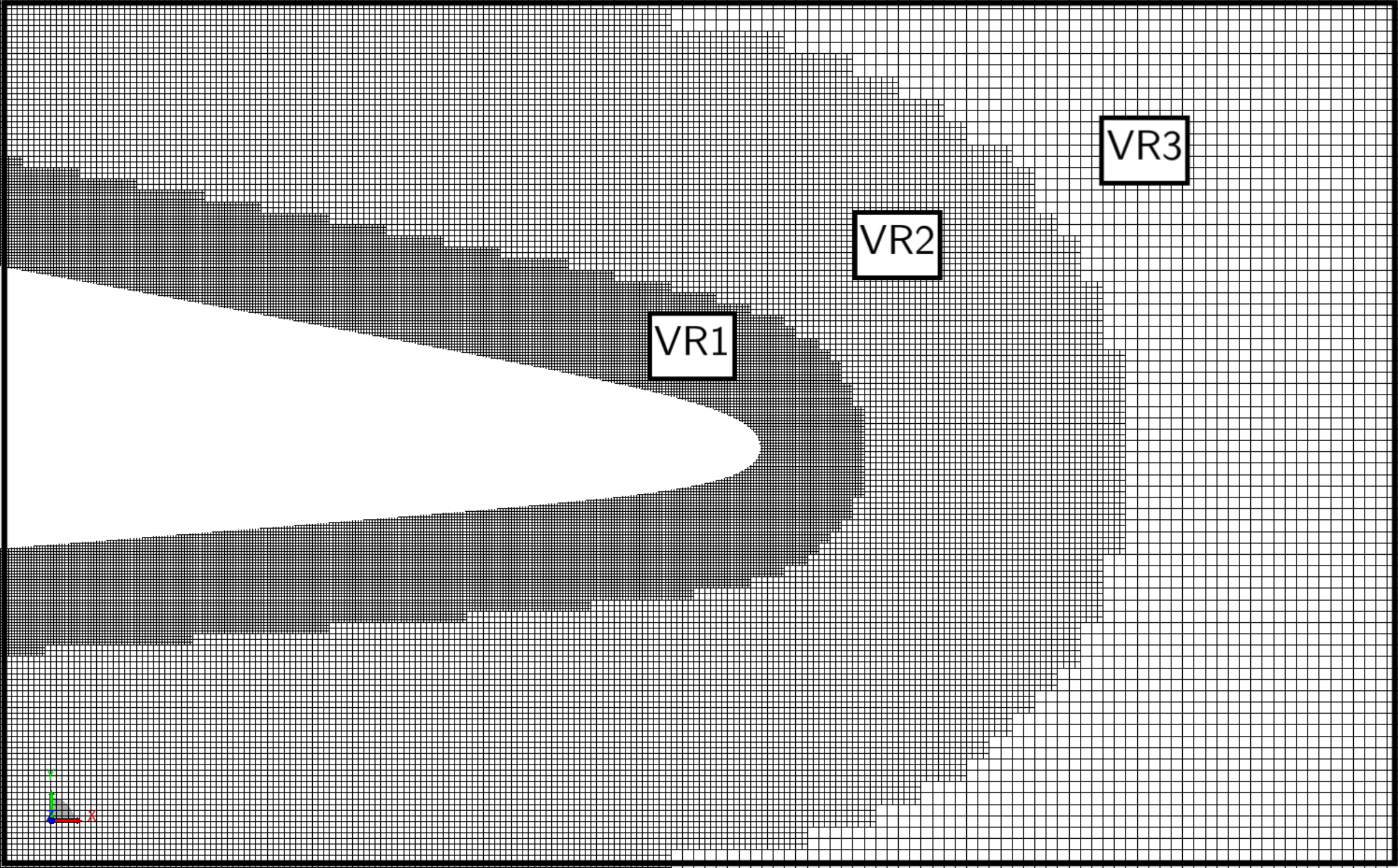}
    \caption{Baseline mesh with extended offset regions.}
    \label{f:baseline1uniform}
\end{subfigure}
\caption{Visualization of different meshes along the trailing edge region.  The first 3 VRs displayed consist of the offset regions around the airfoil.}
\label{meshes}
\end{figure}

A mesh refinement analysis is performed employing four meshes composed of 11 variable resolution (VR) regions that control the voxel size from the airfoil surface to the outer domain. The difference across meshes is only related to the number of voxels along the airfoil chord. The coarse, baseline and fine meshes have approximately $4000$, $5120$, and $6400$ voxels around the airfoil. The three finest VR regions are applied as an offset of the NACA0012 profile (see Figs. \ref{f:coarse1}, \ref{f:baseline1}, and \ref{f:fine1}), while the other 8 VRs are hexahedral boxes enclosing it. The fourth mesh displayed in Fig. \ref{f:baseline1uniform} is generated with the same resolution along the airfoil surface as the baseline mesh. However, it has an extended offset region while maintaining the same voxel size within each VR. This mesh was made specifically for the boundary layer analysis discussed later in this manuscript. The height of each offset is equivalent to $20$ voxels in all meshes, with exception of the fourth mesh, where $35$, $45$, and $55$ voxels are applied in the VR1, VR2, and VR3, respectively. The finest voxel sizes are $0.0246\%$, $0.0195\%$, and $0.0156\%$ of the chord from the coarsest to the finest resolution. The mesh is constructed so that the boundary layer fits inside the two finest offset regions and there is a total of $450$, $512$, and $652$ million voxels in the computational domain from the coarse to the fine grid refinement. Lastly, the mesh in Fig. \ref{f:baseline1uniform} has a total of $1.1$ billion voxels.
\FloatBarrier

To obtain convergence of the flow statistics, the LBM simulations are performed for $115$ convective time units ($t^* = tU_{\infty}/c = 115$) using approximately 700 AMD EPYC 7443 cores for around $2.5 \times 10^5$ CPU hours with the baseline mesh. The NS simulation was performed for $t^* = 105$ (the first $t^* = 30$ were discarded to avoid transient effects), with similar computational resources. However, the LBM calculations required around 10\% of the computational cost of the NS approach.
In the LBM, the initial transient regime is also removed, corresponding to the first $t^* = 40$. With the goal of verifying if the meshes utilized in the data analysis are properly designed, the $y^+$ value for the baseline mesh is calculated and a value of $y^+ \approx 0.6$ is obtained in a region close to the TE, where flow transition occurs.

\section{Results}

Results of the LBM approach are first validated against a wall-resolved LES computed for the compressible NS equations. Comparisons are presented in terms of mean and RMS quantities computed along the flow field and airfoil surface. Then, a vortex dynamics analysis is presented including the effects of intermittency on the flow dynamics and noise generation, besides its consequences on signal processing via Fourier and wavelet decompositions.

\subsection{Mean flow field and surface data}

\begin{figure}[h!]
\centering
   	\begin{subfigure}{0.49\textwidth}
        \includegraphics[trim={0 0.5cm 2cm 0},clip, width=0.943\linewidth]{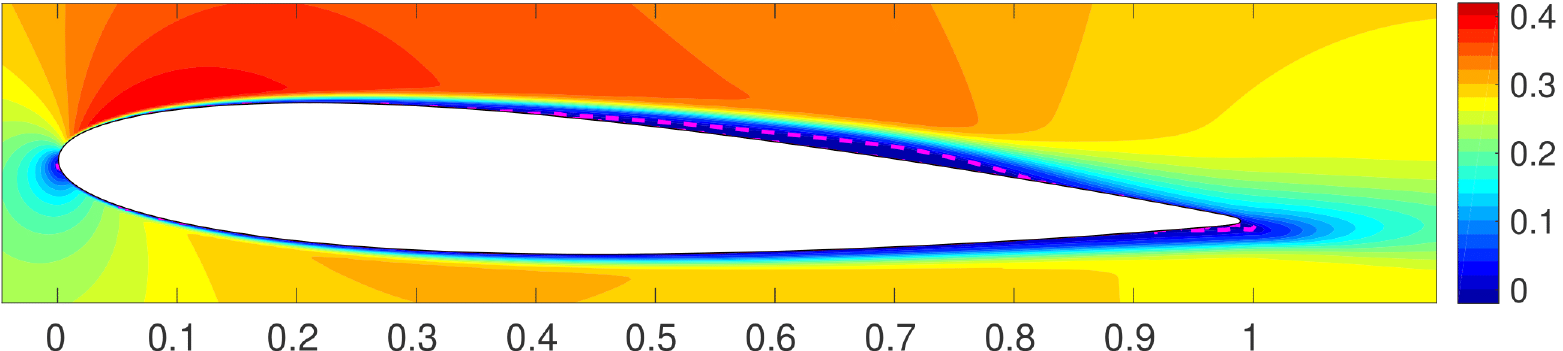}
        \caption{LBM - Coarse mesh}
        \label{f:lbmCoarse}	
 	\end{subfigure}
   	\begin{subfigure}{0.49\textwidth}
        \includegraphics[trim={0 0.5cm 2cm 0},clip, width=0.943\linewidth]{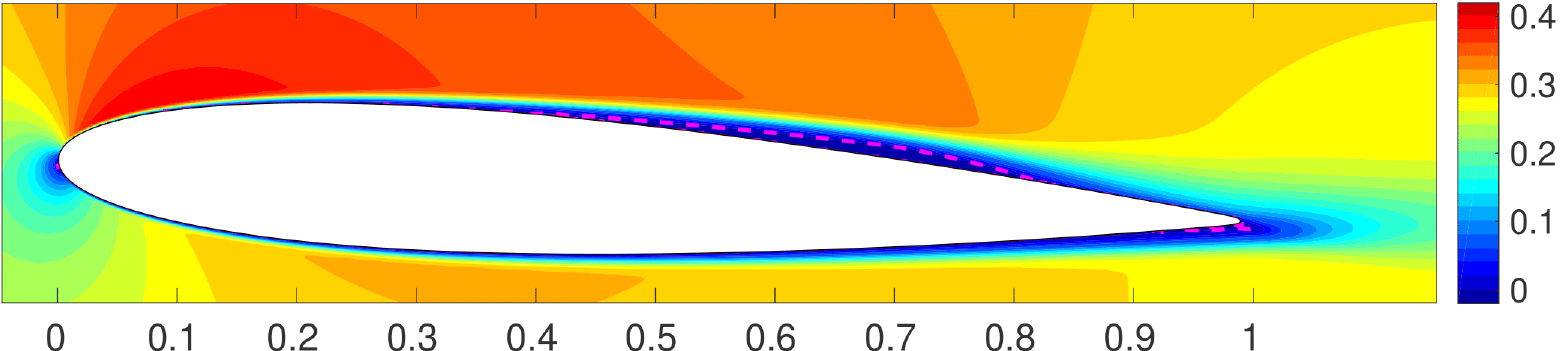}
 	      \caption{LBM - Baseline mesh}
 	      \label{f:lbmBaseline}
 	\end{subfigure}   
   	\begin{subfigure}{0.49\textwidth}
        \includegraphics[trim={0 0.5cm 1.99cm 0},clip, width=0.943\linewidth]{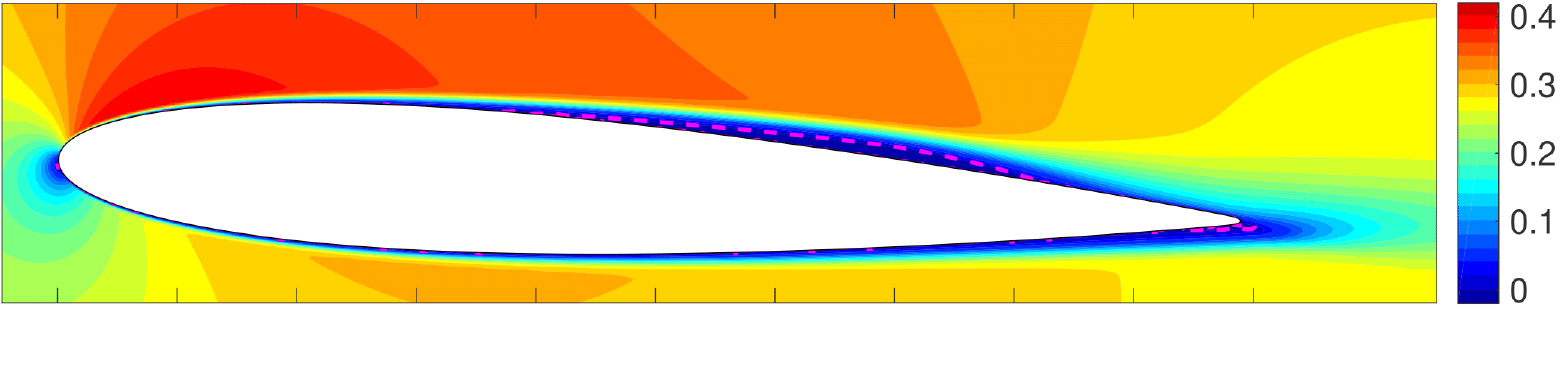}
 	      \caption{LBM - Fine mesh}
 	      \label{f:lbmFine}
 	\end{subfigure}   
   	\begin{subfigure}{0.49\textwidth}
 	\centering
        \includegraphics[trim={0.2cm 0cm 0cm 0cm},clip, width=1\linewidth]{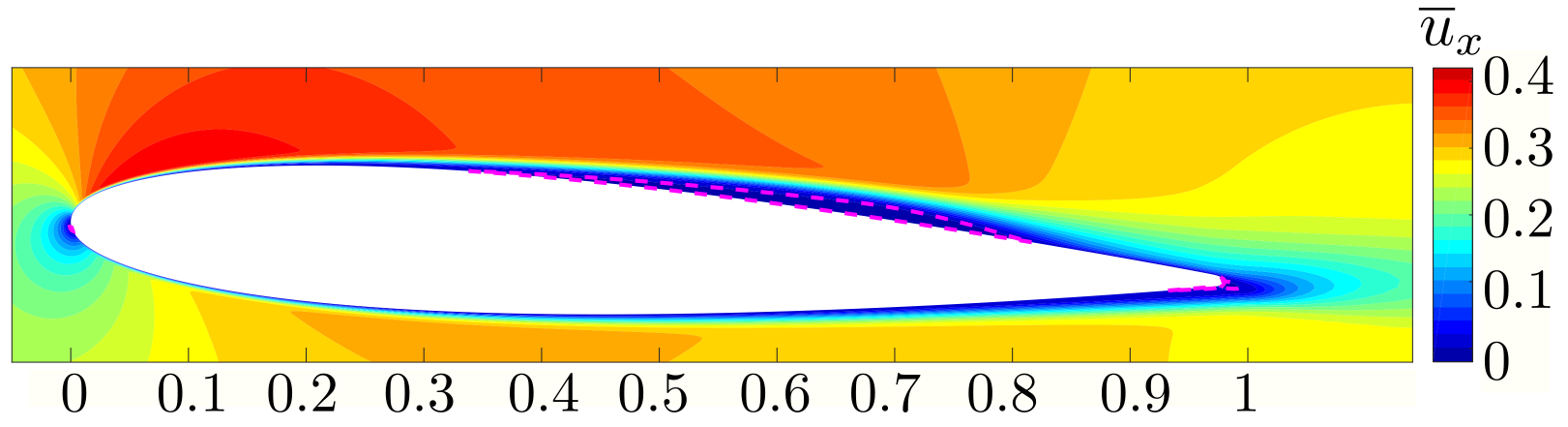}
 	      \caption{NS}
 	      \label{f:NSdata}
 	\end{subfigure}     
\caption{Contours of mean streamwise velocity, $\overline{u}_x$, normalized by the freestream speed of sound (Mach number contours). The magenta dashed lines highlight the reversed flow boundaries which display a separation bubble on the suction side in addition to a small bubble on the pressure side, near the trailing edge. Figure \ref{f:NSdata} is generated with data from Ricciardi et al. \cite{ricciardiJFM} and the same contour levels are use in all plots.}
\label{f:timeAVG} 
\end{figure}

First, we analyze the LBM mean flow field solutions in terms of the LSB size and location. For this, the time- and spanwise-averaged streamwise velocity field, $\overline{u}_x$, is computed and compared against the NS simulation from Ricciardi et al. \cite{ricciardiJFM}, as displayed in Fig. \ref{f:timeAVG}. The magenta dashed lines in both the LBM and NS simulations depict the reversed flow boundaries ($\overline{u}_x < 0$) which include a wide separation bubble on the suction side, besides a small LSB on the pressure side, near the trailing edge. For all LBM simulations, the mean velocity fields display a good agreement with the NS solution.
\FloatBarrier

\begin{figure}[h!]
\centering
   	\begin{subfigure}{0.49\textwidth}
        \includegraphics[trim={0 0.5cm 2.5cm 0},clip, width=0.91\linewidth]{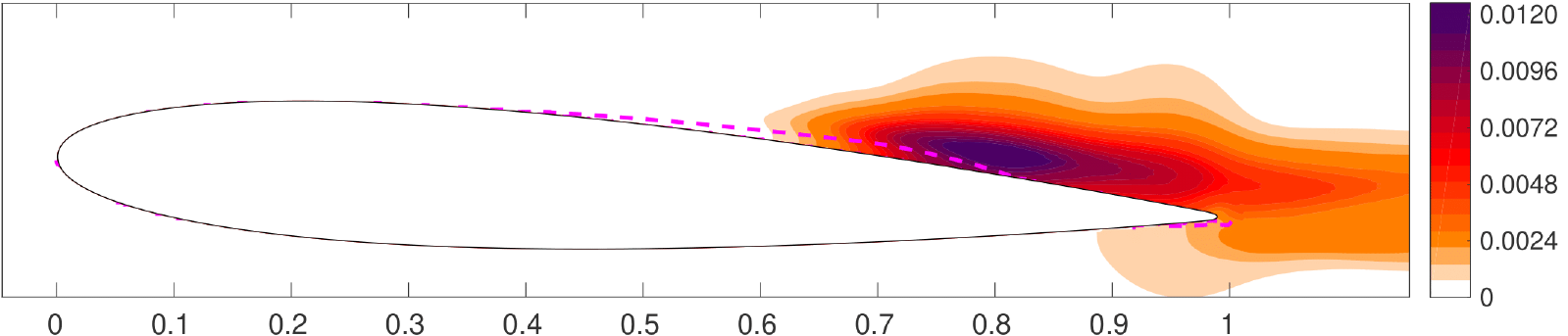}
        \caption{LBM - Coarse mesh}
        \label{f:lbmRMSpcoarse}	
 	\end{subfigure}
   	\begin{subfigure}{0.49\textwidth}
        \includegraphics[trim={0 0.5cm 2.5cm 0},clip, width=0.91\linewidth]{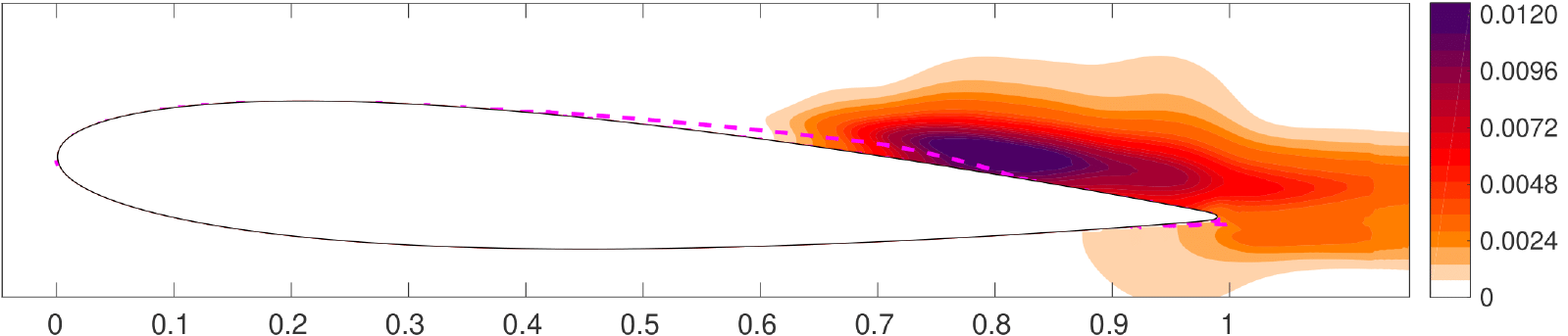}
 	      \caption{LBM - Baseline mesh}
 	      \label{f:lbmRMSpbaseline}
 	\end{subfigure}   
   	\begin{subfigure}{0.49\textwidth}
        \includegraphics[trim={0cm 0cm 2.5cm 0cm},clip, width=0.91\linewidth]{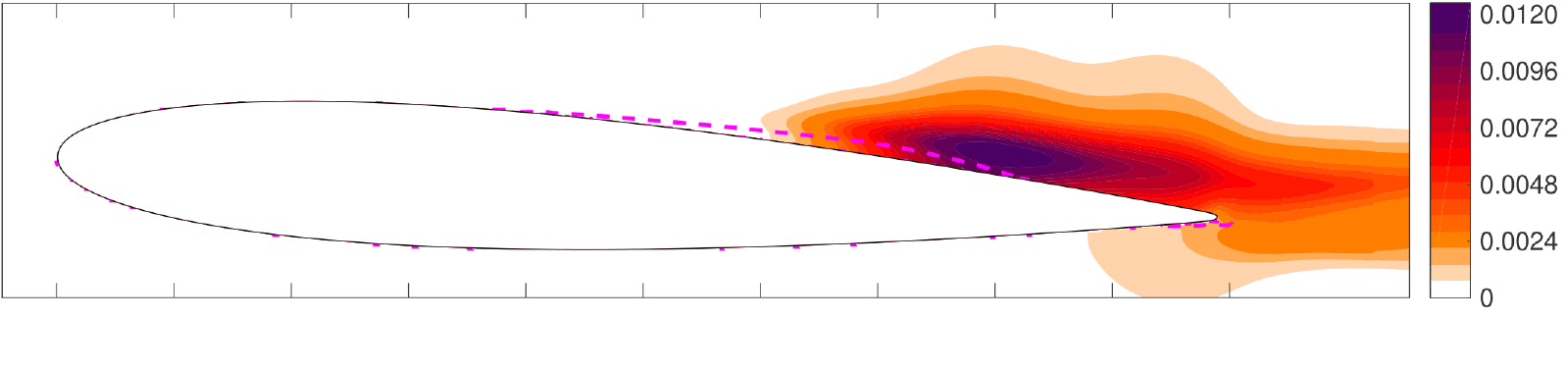}
 	      \caption{LBM - Fine mesh}
 	      \label{f:lbmRMSpfine}
 	\end{subfigure}   
   	\begin{subfigure}{0.49\textwidth}
 	\centering
        \includegraphics[trim={0.2cm 0cm 0cm 0cm},clip, width=1\linewidth]{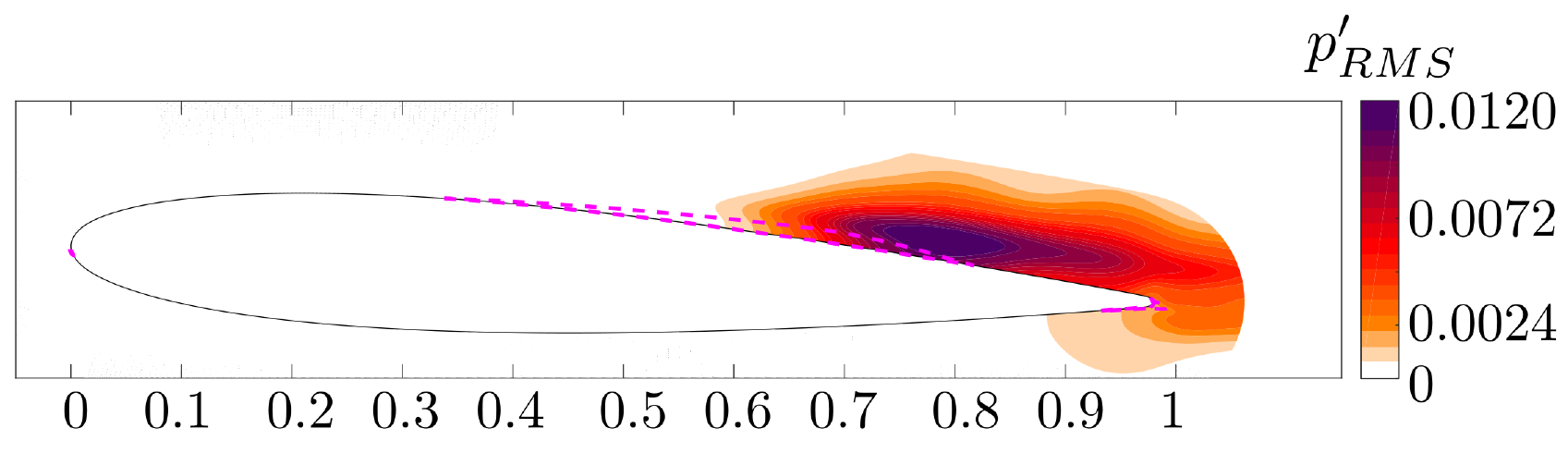}
 	      \caption{NS}
 	      \label{f:cfdRMSp}
 	\end{subfigure}     
\caption{Contours of root-mean-square of pressure fluctuation, $p'_{RMS}$, normalized by the freestream density and speed of sound. The magenta dashed lines highlight the reversed flow boundaries. Figure \ref{f:cfdRMSp} is computed with data from Ricciardi et al. \cite{ricciardiJFM} and the same contour levels are used in all plots.}
\label{f:RMSp} 
\end{figure}

\begin{figure}[h!]
\centering
   	\begin{subfigure}{0.49\textwidth}
        \includegraphics[trim={0 0.5cm 2.1cm 0},clip, width=0.925\linewidth]{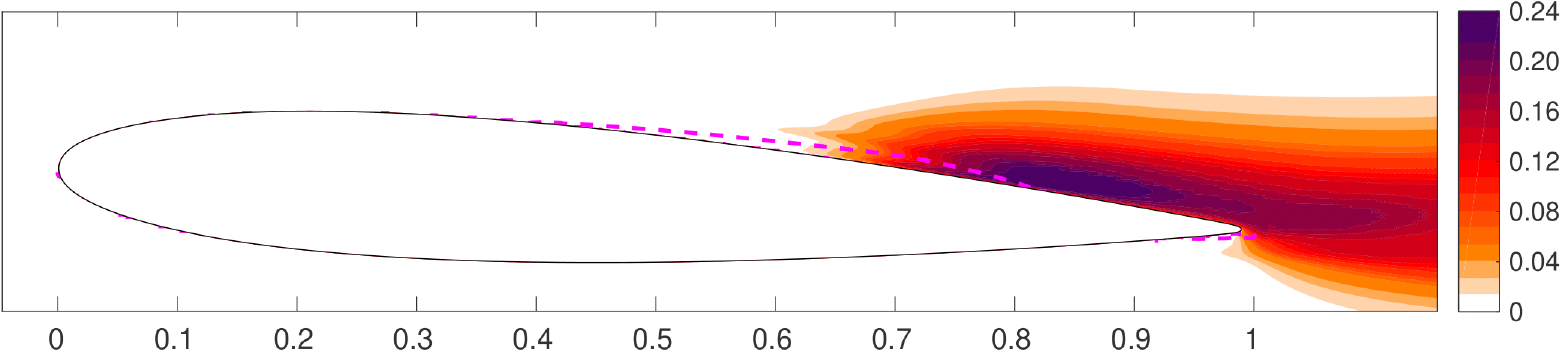}
        \caption{LBM - Coarse mesh}
        \label{f:lbmRMSkcoarse}	
 	\end{subfigure}
   	\begin{subfigure}{0.49\textwidth}
        \includegraphics[trim={0 0.5cm 2.1cm 0},clip, width=0.925\linewidth]{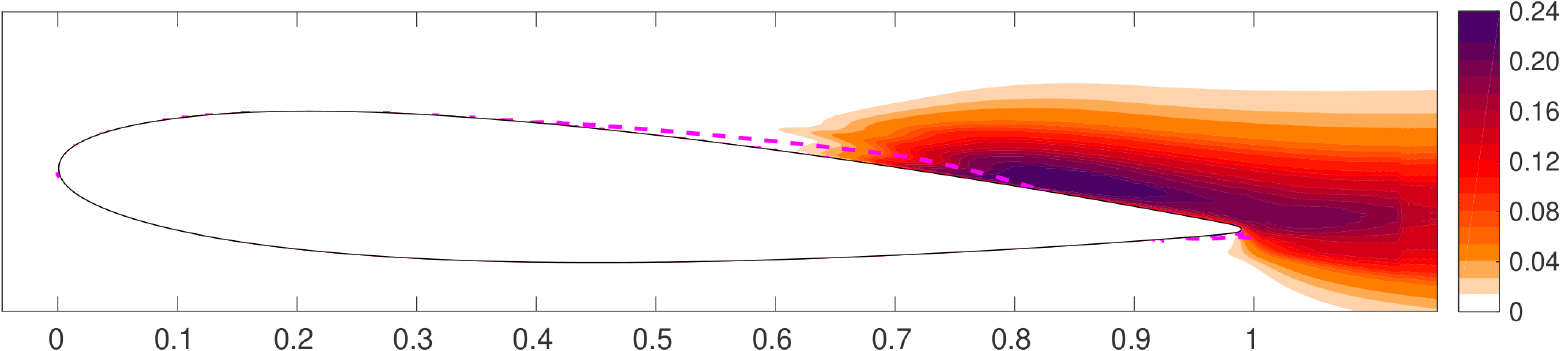}
 	      \caption{LBM - Baseline mesh}
 	      \label{f:lbmRMSkbaseline}
 	\end{subfigure}   
   	\begin{subfigure}{0.49\textwidth}
        \includegraphics[trim={0 0.5cm 2.1cm 0},clip, width=0.925\linewidth]{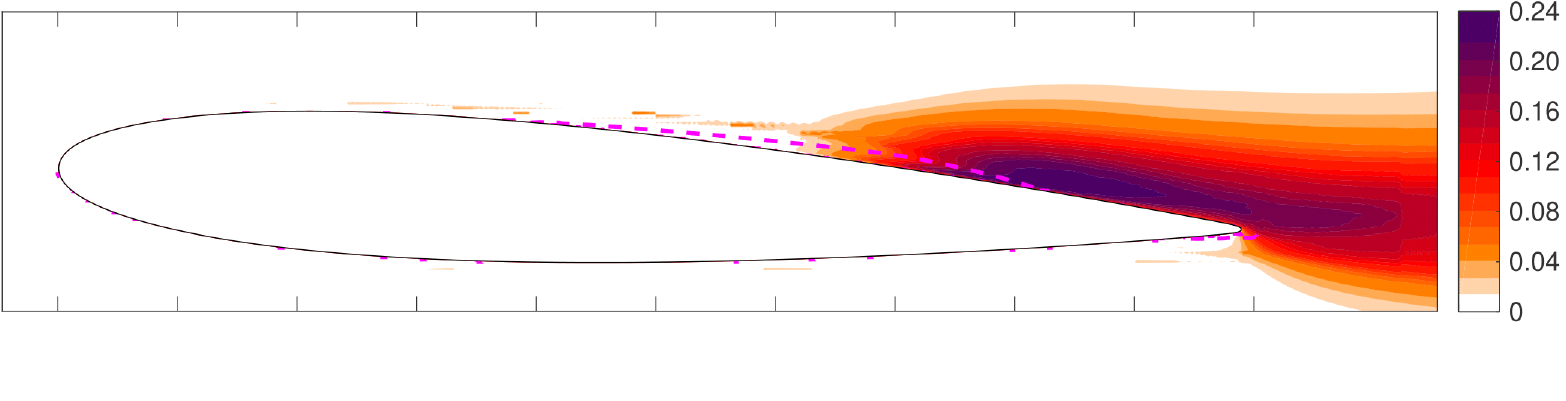}
 	      \caption{LBM - Fine mesh}
 	      \label{f:lbmRMSkfine}
 	\end{subfigure}   
   	\begin{subfigure}{0.49\textwidth}
 	\centering
        \includegraphics[trim={0.2cm 0cm 0cm 0cm},clip, width=1\linewidth]{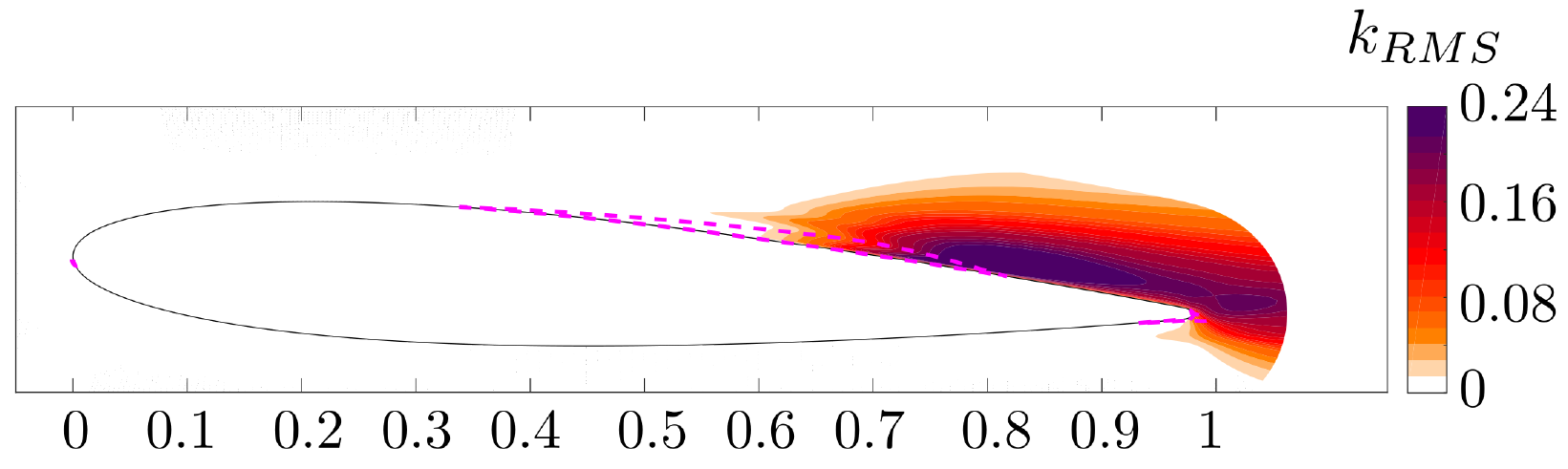}
 	      \caption{NS}
 	      \label{f:cfdRMSk}
 	\end{subfigure}     
\caption{Contours of root-mean-square of the turbulent kinetic energy, $k_{RMS}$, normalized by the freestream velocity. The magenta dashed lines highlight the reversed flow boundaries. Figure \ref{f:cfdRMSk} is computed with data from Ricciardi et al. \cite{ricciardiJFM} and the same contour levels are used in all plots.}
\label{f:RMSk} 
\end{figure}

Figures \ref{f:RMSp} and \ref{f:RMSk} show contours of RMS of the pressure fluctuations ($p'_{RMS}$) and turbulent kinetic energy ($k_{RMS}$), respectively. These quantities are relevant in the context of airfoil self-noise generation since strong pressure fluctuations can lead to high levels of noise radiation due to acoustic scattering at the trailing edge.
They are computed for the coarse, baseline and fine LBM meshes, and for the NS solution. The latter is computed for a small portion of the LES grid. Good agreement is observed between the LBM and NS solutions both in terms of magnitude and location of maximum disturbances, which correspond to the LSB reattachment region on the suction side of the airfoil. These metrics also show the mesh independence since for all LBM cases, the locations and values of highest disturbances are comparable.
\FloatBarrier

\FloatBarrier

\begin{figure}[h!]
    \centering
    \begin{subfigure}[b]{0.49\textwidth}
        \includegraphics[trim={0 0 2cm 0.5cm},clip,width=\linewidth]{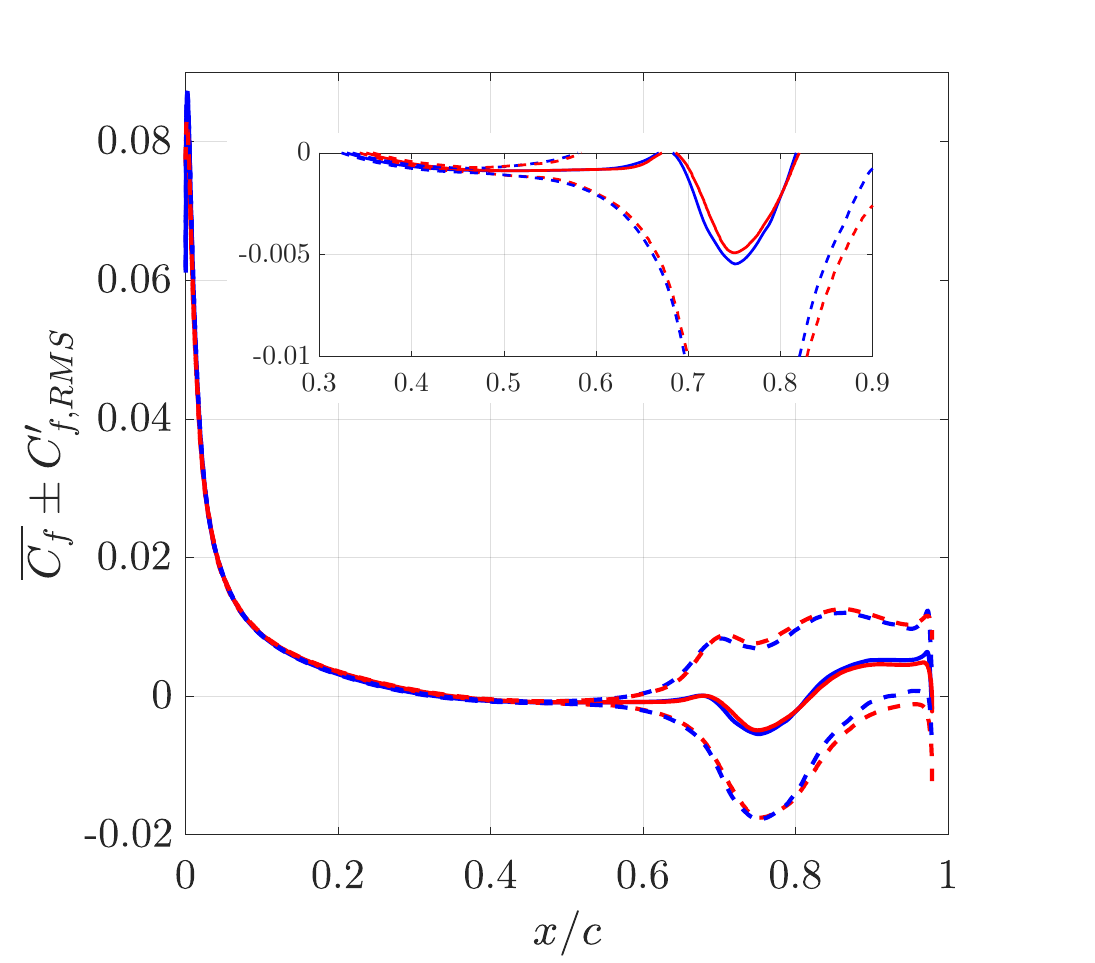}
        \caption{}
        \label{fig:cf1}
    \end{subfigure}
    \hfill
    \begin{subfigure}[b]{0.49\textwidth}
        \includegraphics[trim={0 0 2cm 0.5cm},clip,width=\linewidth]{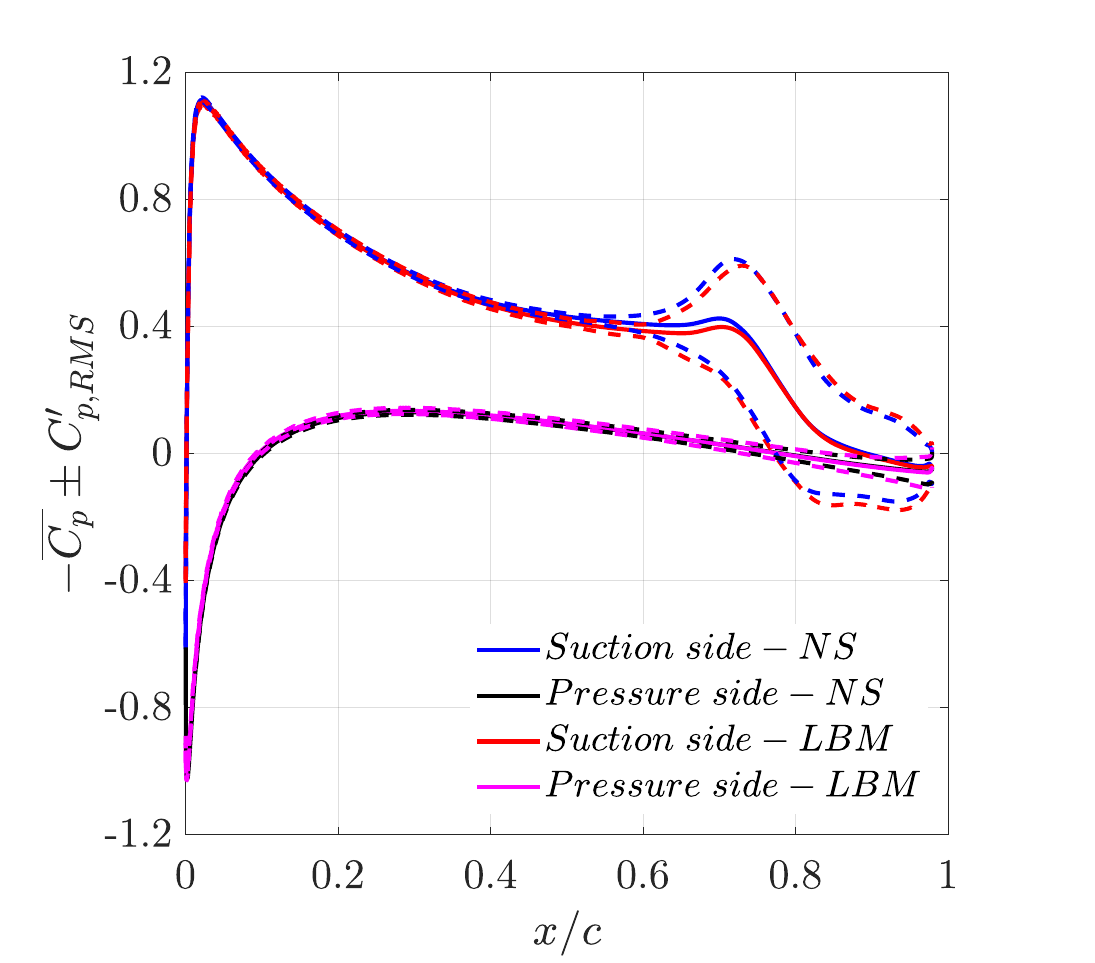}
        \caption{}
        \label{fig:cp1}
    \end{subfigure}
\FloatBarrier

    \caption{Mean and RMS distributions of (a) suction side skin friction coefficient and (b) pressure coefficient along the airfoil surface. The solid lines display the mean values while the dashed ones display the mean $\pm$ the RMS values.}
    \label{fig:surf}
\end{figure}

The previous flow field comparisons indicate that the LBM solutions are mesh independent. Here, more quantitative analyses are provided by inspection of the skin friction and pressure coefficients. 
The present comparisons are shown using the LBM baseline solution and the NS results. 
The mean (time- and spanwise-averaged) skin friction coefficient, $\overline{C_f}$, is plotted with solid lines in Fig. \ref{fig:cf1} over the airfoil suction side for the NS and LBM approaches. In the same figure, the RMS values $C'_{f,RMS}$ are also added to the mean solution to display the regions with strong flow fluctuations. These results appear with dashed lines. Overall, both the mean and RMS LBM values closely follow the NS results.
The LBM is able to reproduce the double detachment profile observed in the $\overline{C_f}$ plot at around $x/c \approx 0.35$ and $x/c \approx 0.68$. While the first detachment region is due to the separation bubble, the second is related to strong vortical structures that are shed from the shear layer induced by the bubble. The figure also shows that the skin friction fluctuations are intense on the airfoil suction side downstream of the bubble, up to the trailing edge. 

The mean pressure coefficient $-\overline{C_p}$ is shown in Fig. \ref{fig:cp1} by solid lines. Pressure fluctuations $C'_{p,RMS}$ are also added to the mean values, being depicted by the dashed lines. Due to the adverse pressure gradient, a gentle drop of $-\overline{C_p}$ is observed on the suction side from a region near the leading edge until $x/c \approx0.7$. Then, a sharper pressure drop occurs due to vortex shedding from the separation bubble. This effect can be seen in the region with high pressure fluctuations $C'_{p,RMS}$ from $x/c \approx 0.6$ until the trailing edge. The mean coefficient and RMS pressure values maintain the same trends and magnitudes when compared to the wall-resolved LES.
%
\FloatBarrier

\subsection{Velocity profiles}

\begin{figure}[h!]
    \centering
    \begin{subfigure}[b]{0.9\textwidth}
        \includegraphics[trim={0.2cm 0 0 0},clip,width=\linewidth]{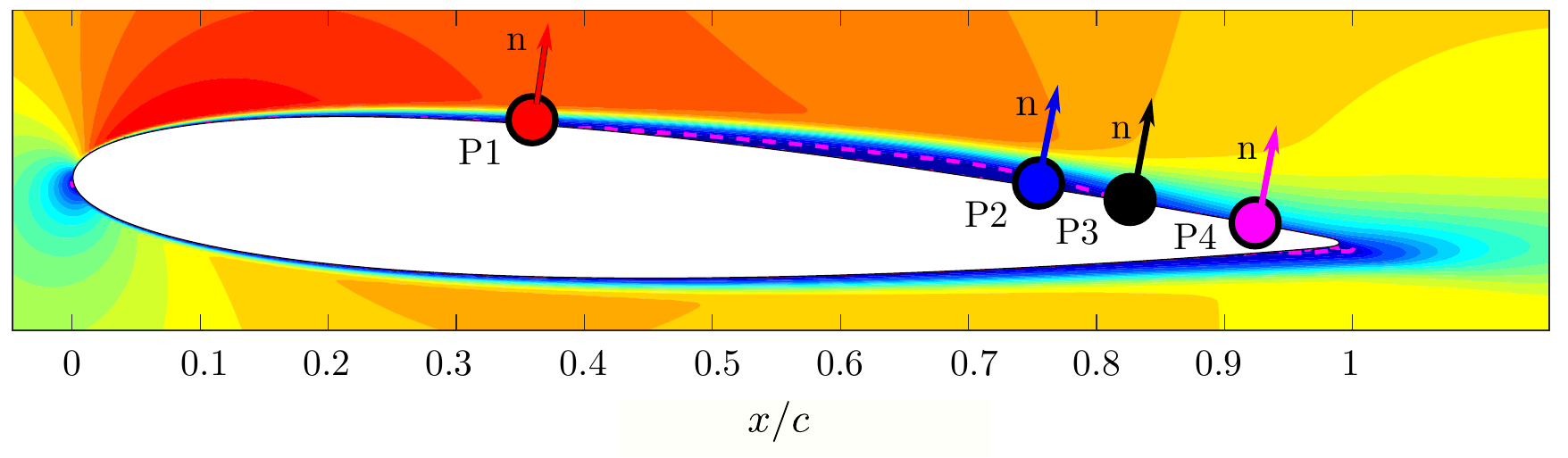}
        \caption{}
        \label{fig:points}
    \end{subfigure}

    \begin{subfigure}[b]{0.49\textwidth}
        \includegraphics[trim={0 0 0cm 0},clip,width=\linewidth]{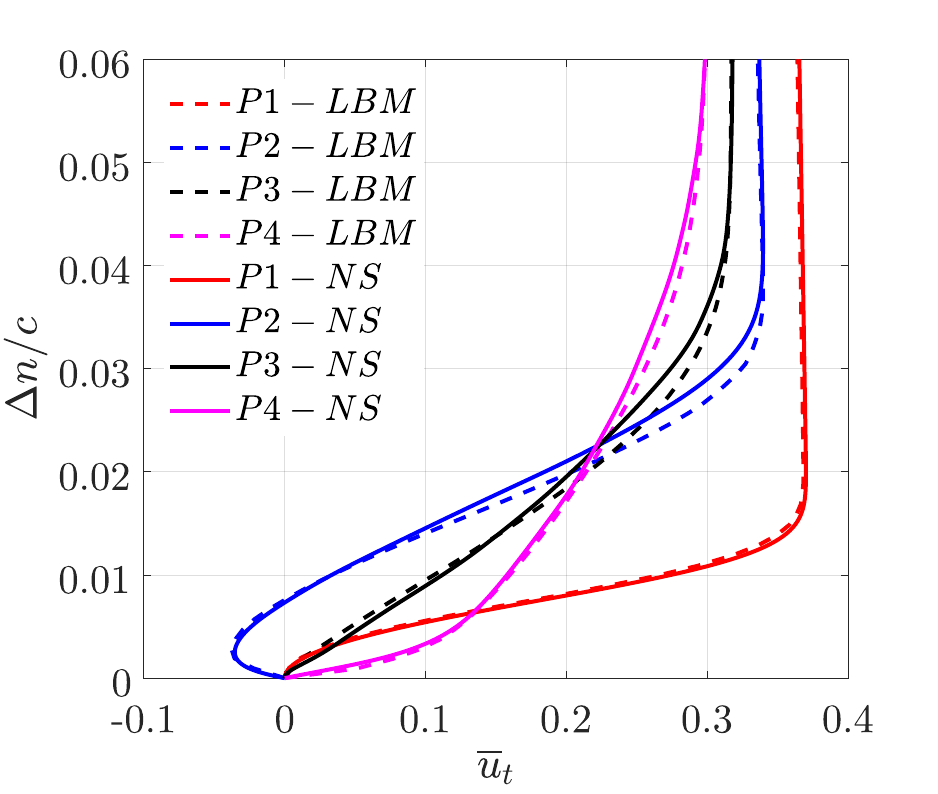}
        \caption{}
        \label{fig:utavg}
    \end{subfigure}
    \begin{subfigure}[b]{0.49\textwidth}
        \includegraphics[trim={0 0 0cm 0},clip,width=\linewidth]{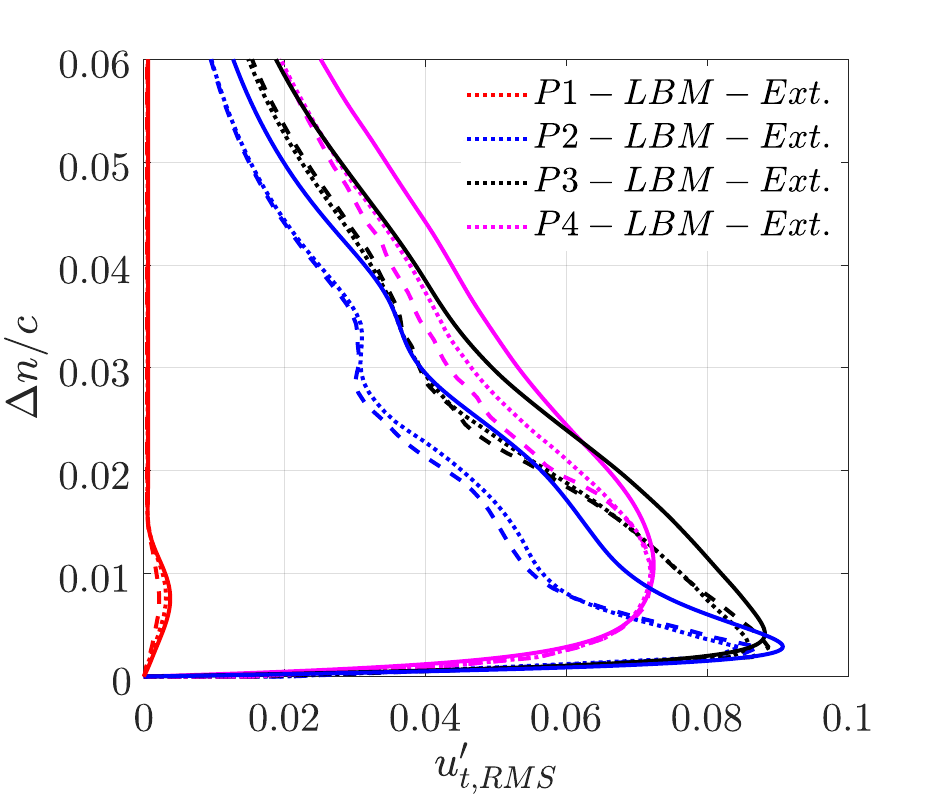}
        \caption{}
        \label{fig:utrms}
    \end{subfigure}

    \caption{Analysis of velocity profiles at four locations over the airfoil suction side. (a) P1 corresponds to the boundary layer separation point; P2 has the highest magnitude of $\overline{C_f}$; P3 matches the boundary layer reattachment point; and P4 is $10\%$ downstream from P3. (b) Mean tangential velocity, $\overline{u}_t$, profiles. (c) Tangential velocity fluctuation, $u'_{t,RMS}$, profiles. The dashed lines are computed for the LBM baseline mesh, while the dotted lines correspond to the mesh with extended offset regions near the wall.}
    \label{fig:surf}
\end{figure}

In this subsection, we analyze the evolution of the boundary layer velocity profiles over the suction side.
Profiles are extracted at four positions displayed in Fig. \ref{fig:points}. These locations correspond to the point of boundary layer (BL) separation (red); the point of maximum skin friction coefficient within the region of high $C'_{f,RMS}$ (blue); the reattachment point (black); and a position $10\%$ downstream from the BL reattachment (magenta). The mean tangential velocity, $\overline{u}_t$, profiles extracted along the wall-normal direction $\Delta n/c$ are displayed in Fig. \ref{fig:utavg}. 
Overall, the mean velocity profiles show a good agreement with the LES results both in terms of magnitude as well as shape. Here, results are shown only for the baseline mesh. However, mean velocity profiles computed for all LBM meshes presented similar comparisons to the NS solution.
The red line indicates a zero wall-normal derivative in $\overline{u}_t$, as expected for the BL detachment point. The blue line shows a maximum reversed flow of $-0.12U_{\infty}$, closely matching the value from the LES. Near the wall, the magenta velocity profile is fuller than the black one since it is computed in a region where transition to turbulence occurs. However, since the adverse pressure gradient increases downstream, the velocity profile of the black line depicts a higher velocity away from the wall.

Profiles of tangential velocity fluctuations, $u'_{t,RMS}$, are shown in Fig. \ref{fig:utrms}. The solid lines depict the NS solutions, while the dashed and dotted lines are computed for the baseline mesh (Fig. \ref{f:baseline1}) and that with the extended offsets (Fig. \ref{f:baseline1uniform}), respectively. As expected, lower velocity fluctuation levels are observed at the separation region. However, the fluctuations increase considerably downstream due to vortex shedding and transition to turbulence, as will be discussed later. A comparison with the NS solution shows similar trends, especially close to the airfoil surface. The RMS values of the LBM solutions deviate from the NS ones at wall-normal distances higher than $\Delta n/c > 0.005$. In order to investigate this difference, solutions computed by the mesh with the extended offset regions are also plotted with dotted lines. It is noticed that even with this more refined mesh, the RMS values are comparable with those from the regular baseline grid. Therefore, this difference could be justified by the high sensitivity of this metric with respect to variations in the chord position rather than a mesh refinement implication. 

Although small differences in the boundary layer mean and RMS velocity profiles are observed between the LBM and NS solutions, the results show that the former approach is able to capture the physics of the present transitional flow. 
This shows that the DNS capability of the LBM can be considered as an alternative approach to numerical simulations of transitional flows, and a promising solution for flows around complex geometries.

\FloatBarrier



\subsection{Vortex dynamics} \label{s:vortex}

\begin{figure}[h!]
\centering
   	\begin{subfigure}{0.44\textwidth}
 	\centering
       	\includegraphics[width=1\linewidth]{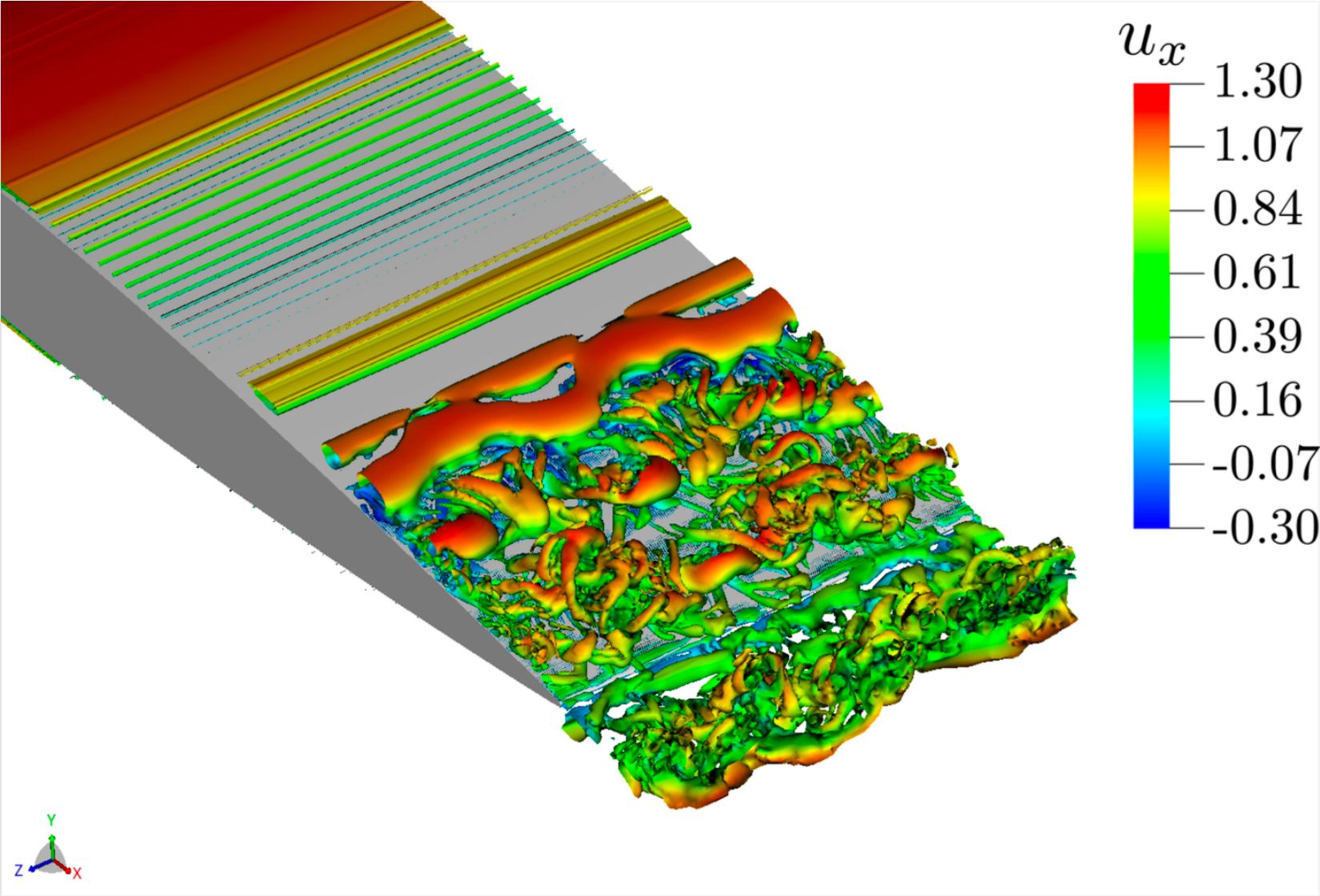}
 		\caption{Turbulent structures at the TE}
 	      \label{f:lbmS}
 	\end{subfigure}
   	\begin{subfigure}{0.44\textwidth}
 	\centering 
       	\includegraphics[width=1\linewidth]{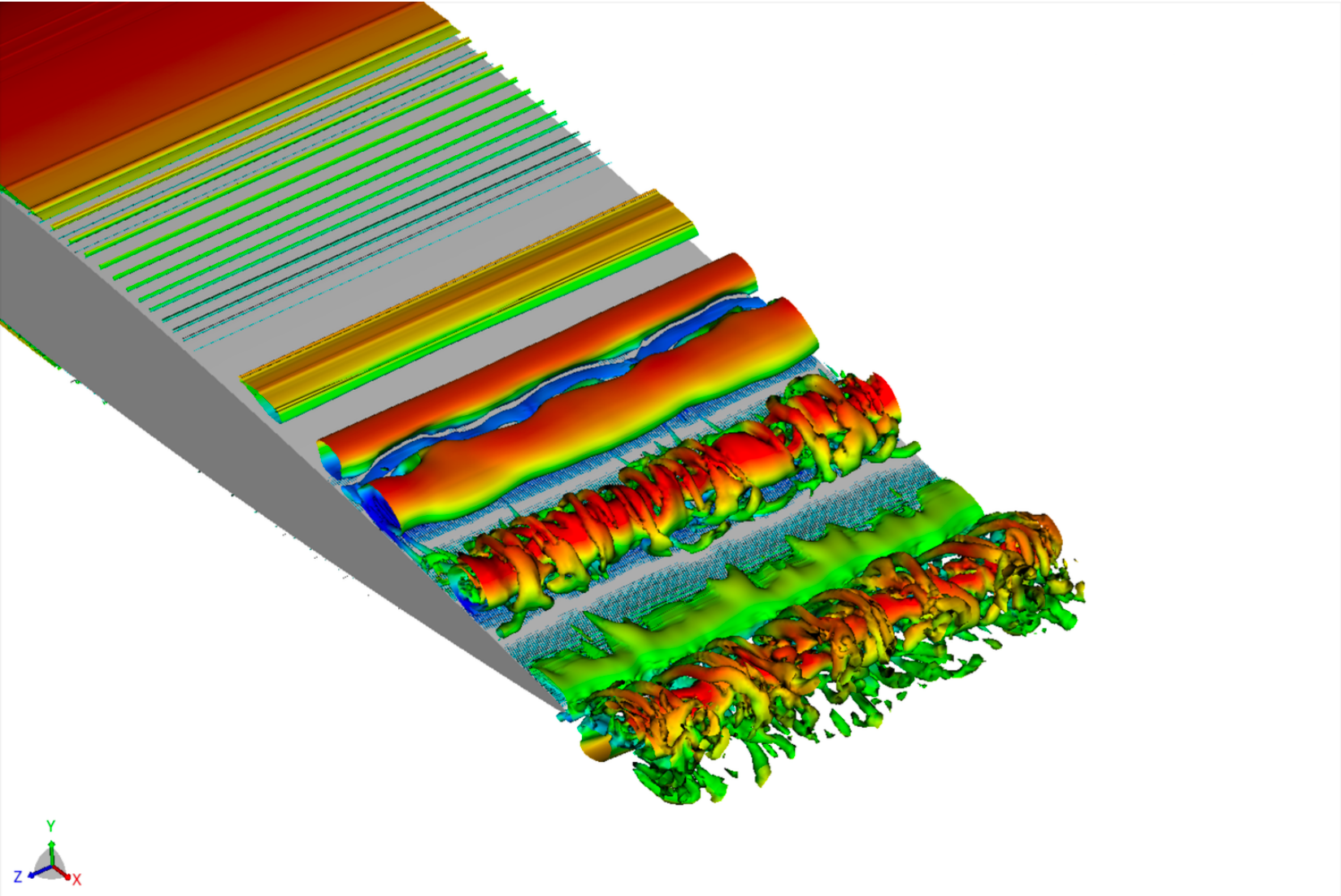}
 		\caption{Spanwise coherent structures at the TE}
 	      \label{f:lbmV}
 	\end{subfigure}    
\caption{Isosurfaces of $\lambda_2$-criterion colored by the instantaneous streamwise velocity $u_x$. Two regimes are shown where (a) shows  smaller-scale turbulent eddies with lower coherence, and (b) displays spanwise-correlated structures shed from the bubble.}
\label{f:cfdvslbm} 
\end{figure}

In this section, we investigate the vortex dynamics of the present flow. Although the mean flow fields and the boundary layer analysis provide a comparison between the LBM and NS methodologies, the study of the vortex dynamics reveals further details of the unsteady flow features. Ricciardi et al. \cite{ricciardiJFM} show that vortex pairing and merging may lead to coherent structures transported through the trailing edge. In some cases, the pairing may not be successful and lead to bursting of fine turbulent scales. Results shown in Fig. \ref{f:cfdvslbm} compare snapshots with isosurfaces of 
$\lambda_2$ criterion colored by the instantaneous streamwise velocity, $u_x$, normalized by $U_\infty$. While the left plot shows uncorrelated turbulent structures at the trailing edge, the right one shows the transport of spanwise coherent vortices shed by the bubble. Hence, the LBM is able to detect the intermittent laminar–turbulent transition that affects the convection of coherent structures from the laminar separation bubble over the suction side toward the TE. 


\FloatBarrier

\begin{figure}[h!]
\centering
        \includegraphics[width=1\linewidth]{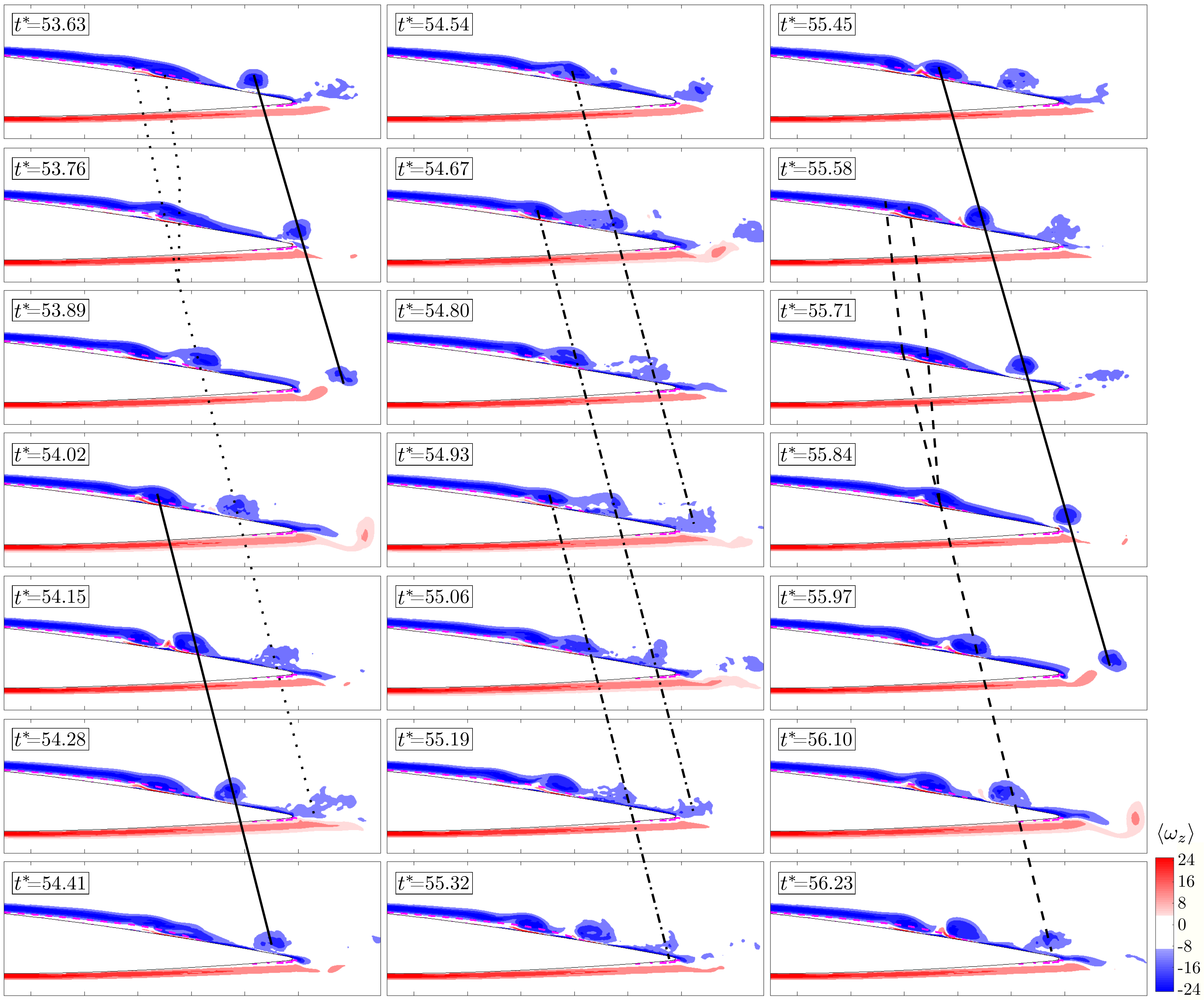}
\caption{Spanwise-averaged z-vorticity flow fields, $\langle {\omega}_z \rangle$, at selected times. The magenta dashed line delimits the reversed mean flow region ($\overline{u}_x < 0$). The various black lines connecting the sub-figures display four distinct flow dynamics: Successful vortex pairing with a coherent structure over the airfoil (- - -); unsuccessful vortex pairing with the formation of turbulent structures (...); transport of a single coherent vortex (---); and a single vortex bursting and transitioning to turbulence (-.-).}
\label{f:vortAVG} 
\end{figure}

To investigate the temporal evolution of the LSB vortex shedding, Fig. \ref{f:vortAVG} displays spanwise-averaged  z-vorticity contours, $\langle \omega_z \rangle$. The simulation time is shown in terms of $t^*$ on the upper left corner of all plots. In this figure, red and blue contours represent positive and negative values of $\langle \omega_z \rangle$, respectively. Higher and more concentrated values of vorticity are associated with higher spanwise coherence, which can be connected to the trailing-edge noise generation.  

As can be seen from Fig. \ref{f:vortAVG}, the flow dynamics is dominated by events on the airfoil suction side, where vortices are shed from the LSB. This effect is typical of low to moderate Reynolds number airfoil flows as discussed in Refs. \cite{probsting2015, ricciardi2022PRF}. Different types of shedding are observed, similarly to the results from Ref. \cite{ricciardiJFM}. First, a single vortex can be formed and shed while maintaining its coherence after reaching the TE. This process is outlined in a sequence of plots with solid black lines. In contrast, a vortex can transition to turbulence and breakdown before reaching the TE as highlighted in the dashed-dotted lines. During the shedding process, there is a possibility that the spanwise coherent vortices undergo a process of vortex pairing. The third flow dynamics observed occurs when there is a successful vortex pairing which creates a coherent structure that reaches the TE. This feature is depicted by dashed lines. Lastly, the breakdown of an unsuccessful vortex pairing is shown by dotted lines. 

\FloatBarrier

\begin{figure}[h!]
\centering
       	\includegraphics[width=0.9\linewidth]{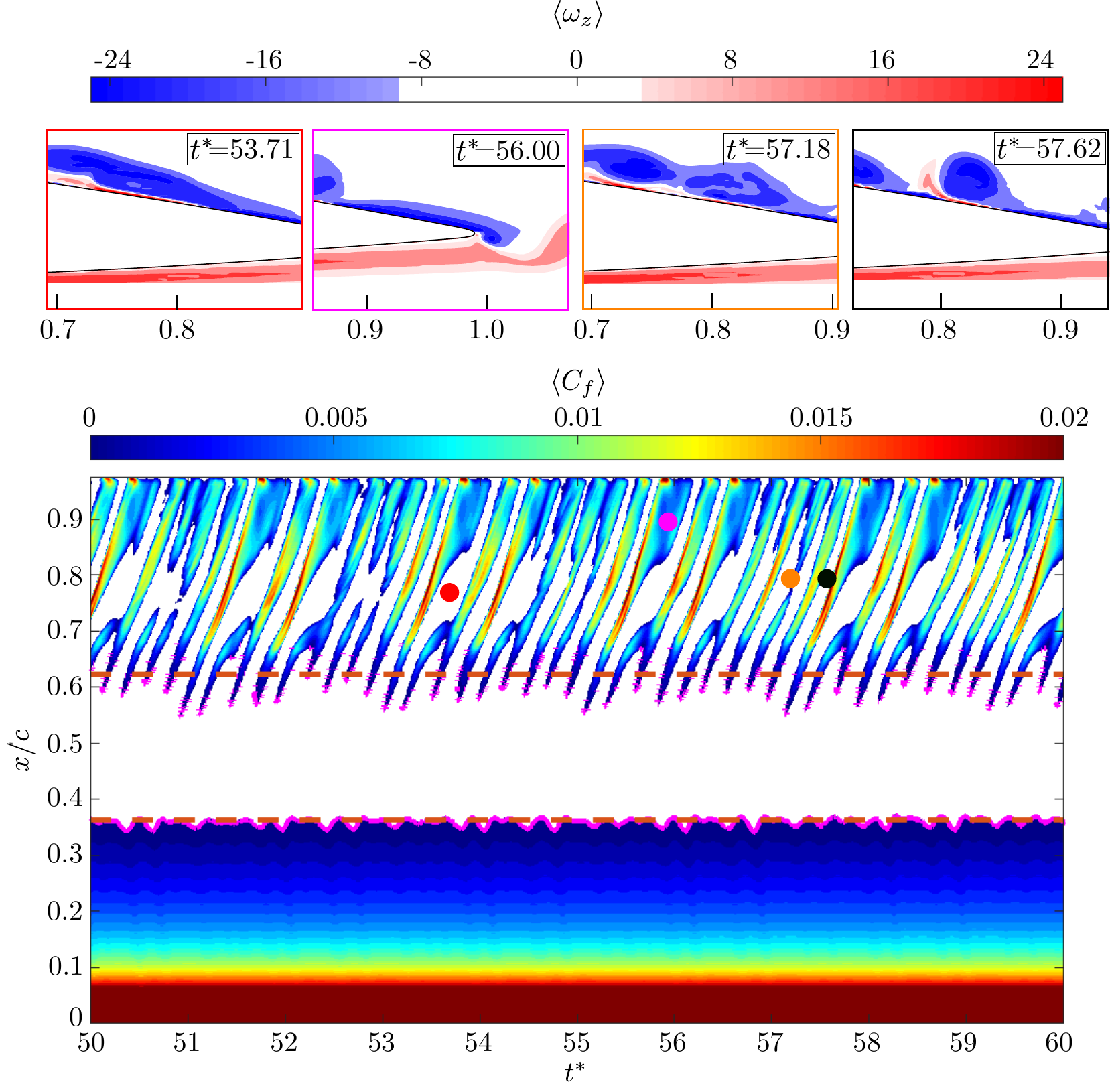}
 	      \label{f:CfVortZ}     
\caption{Instantaneous spanwise-averaged friction coefficient, $\langle C_f \rangle$ along the airfoil suction side. The brown lines depict the mean values of the separation point ($x/c = 0.36$) and reattachment point ($x/c = 0.62$). The top row displays spanwise-averaged z-vorticity, $\langle \omega_z \rangle$, at different instants marked by the colored circles.}
\label{f:Cfanalysis} 
\end{figure}

The vortex dynamics depicted in Fig. \ref{f:vortAVG} can be also analyzed through the spanwise-averaged friction coefficient, $\langle C_f \rangle$, plotted along the airfoil suction side with respect to time, as shown in Fig. \ref{f:Cfanalysis}. This quantity is negatively proportional to $\langle {\omega}_z \rangle$.
For instance, the white regions in Fig. \ref{f:Cfanalysis} represent regions of flow recirculation at the wall ($\langle C_f \rangle <0$ and $\langle {\omega}_z \rangle > 0$). On the other hand, colored regions in the plot are associated with attached flow where $\langle C_f \rangle > 0$ and $\langle {\omega}_z \rangle < 0$ at the wall. Furthermore, the white region where $0.4 \lesssim x/c \lesssim 0.6$ highlights the position of the LSB and the horizontal brown dashed lines at $x/c = 0.36$ and $x/c = 0.62$ delimit the mean position of the LE and TE of the bubble, respectively. These lines are obtained by averaging the position at which $\langle C_f \rangle$ switches sign (magenta markers) between $x/c \approx 0.5$ and $x/c \approx 0.7$, highlighting the bubble TE motion. A magenta line also marks the instantaneous position of the bubble LE.  

As can be seen from the figure, the bubble depicts a breathing motion, where the  excursions of the reattachment location are much larger than those from the separation point. 
Thus, the reattachment point of the bubble has a major role in the intermittency and dynamics of the coherent structures shed from the LSB. Stripes with varying colors of $\langle C_f \rangle$ are observed and their magnitude and shape can be related with the type of intermittent event that occurs over the airfoil suction side. To illustrate this, different colored points are marked in Fig. \ref{f:Cfanalysis} to highlight the different instantaneous flow features.
A red circle is placed at $t^* = 53.71$ and it depicts the instant when a vortex pairing is observed in the above snapshot of $\langle {\omega}_z \rangle$. Although the coherent structure is characterized by a negative z-vorticity, it is possible to observe that a thin layer of positive z-vorticity forms under the pairing. This thin layer is responsible for the thick white region around the red circle in the $\langle C_f \rangle$ figure. The positive vorticity at the wall is induced by the opposite sign vorticity originated by the vortex pairing. Wang et al. \cite{wang2016} experimentally demonstrated the same vorticity induction in a vortex merger in ground proximity. 

A wide region of positive $\langle C_f \rangle$ (negative vorticity at the wall) is depicted at $t^*=56.00$ (magenta circle), representing an instant when an attached  boundary layer forms near the trailing edge. Single vortex breakdown/bursting is translated to a split in the friction coefficient due to the positive/negative alternate z-vorticity layers at the wall, as displayed in $t^*=57.18$ (orange circle). At $t^*=57.62$ (black circle), a strong and coherent vortex is formed  inducing a positive z-vorticity at the wall. The coherent vortex also enables a high positive $\langle C_f \rangle$ region next to the flow separation, possibly due to another vorticity induction from positive to negative. Lastly, the inclination of the stripes representing the intermittent events are directly related to the convective velocity of the structures over the airfoil and, although the flow dynamics are distinct, the angle of $\langle C_f \rangle$ with respect to the x-axis (time) are similar across all events towards the TE.

\FloatBarrier

\begin{figure}[h!]
\centering
       	\includegraphics[width=1\linewidth]{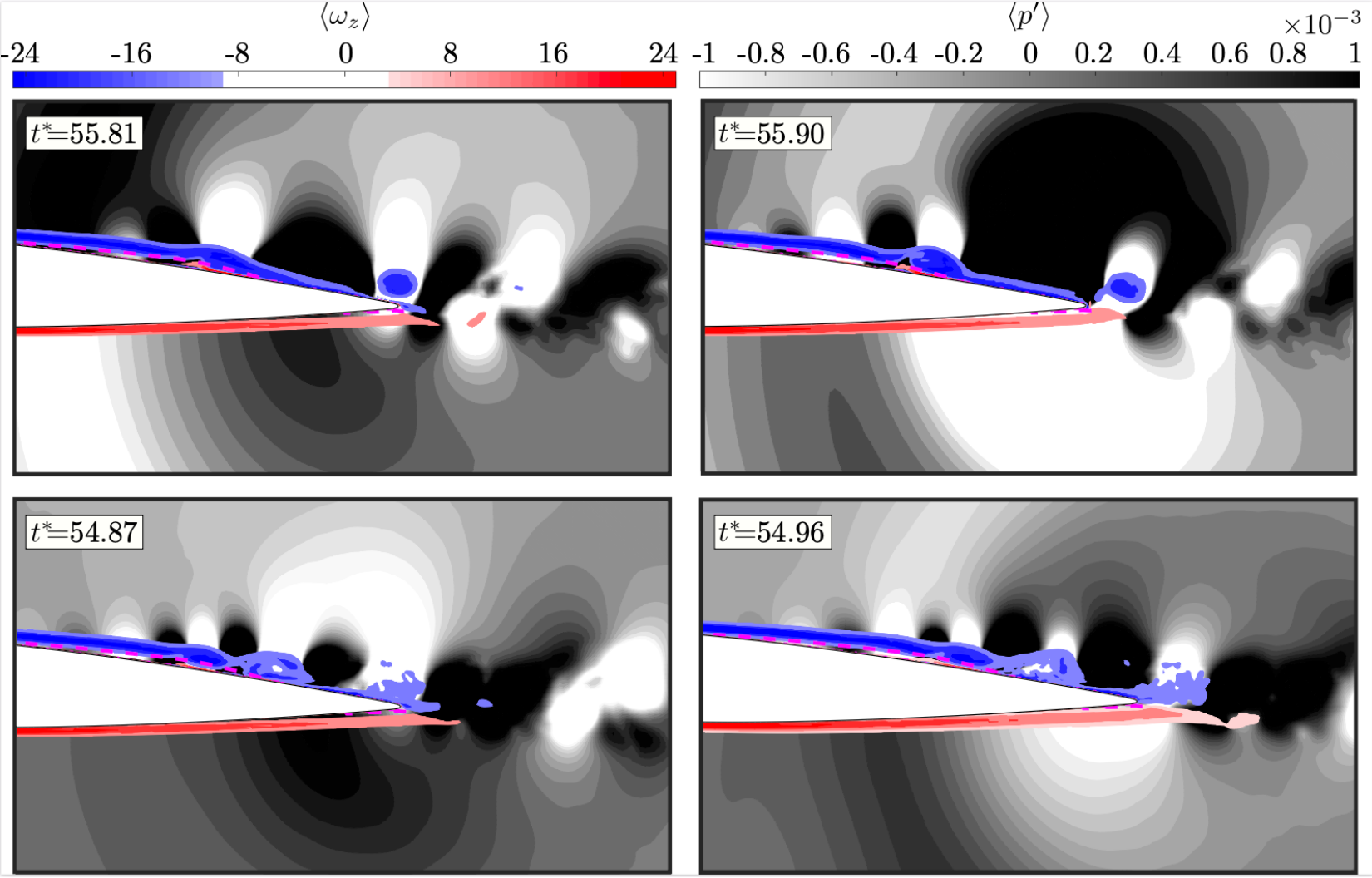}
 	      \label{f:acoustic}    
\caption{Instantaneous contours of spanwise-averaged z-vorticity $\langle \omega_z \rangle$ (color) overlaid on the spanwise-averaged pressure fluctuation $\langle p' \rangle$ (gray) displaying the acoustic emission from coherent (top row) and uncorrelated (bottom row) structures that reach the trailing edge. The left column shows the structures above the TE, while the right column displays the pulses due to acoustic scattering.}
\label{f:acfields} 
\end{figure}

Shedding of spanwise-coherent structures at the trailing edge of the airfoil lead to intense acoustic scattering as discussed in Refs. \cite{Ffowcs1970, SanoPRF2019}. In the present flow, the intermittent vortex dynamics observed in Figs. \ref{f:vortAVG} and \ref {f:Cfanalysis} is responsible for airfoil self-noise generation at different frequencies and amplitudes. This phenomenon is displayed in Fig. \ref{f:acfields}, which depicts instants when a spanwise-coherent vortex reaches the TE (top row) and when a less-correlated structure is transported along the TE (bottom row). Results are shown in terms of $\langle \omega_z \rangle$ (color) and $\langle p' \rangle$ (gray scale) contours. The coherent structure observed in the top-left subfigure is associated with a strong negative hydrodynamic pressure fluctuation which scatters at the trailing edge leading to a strong black-and-white dipolar acoustic pulse on the airfoil, observed in the top-right subfigure. In contrast, the less-correlated structure in the bottom-left subfigure generates a weaker acoustic pulse shown in the bottom-right plot. 

\FloatBarrier

\subsection{Intermittency} \label{s:interm}

\begin{figure}[h!]
\centering
        \includegraphics[width=1.0\textwidth, trim={50 0 20 0}, clip]{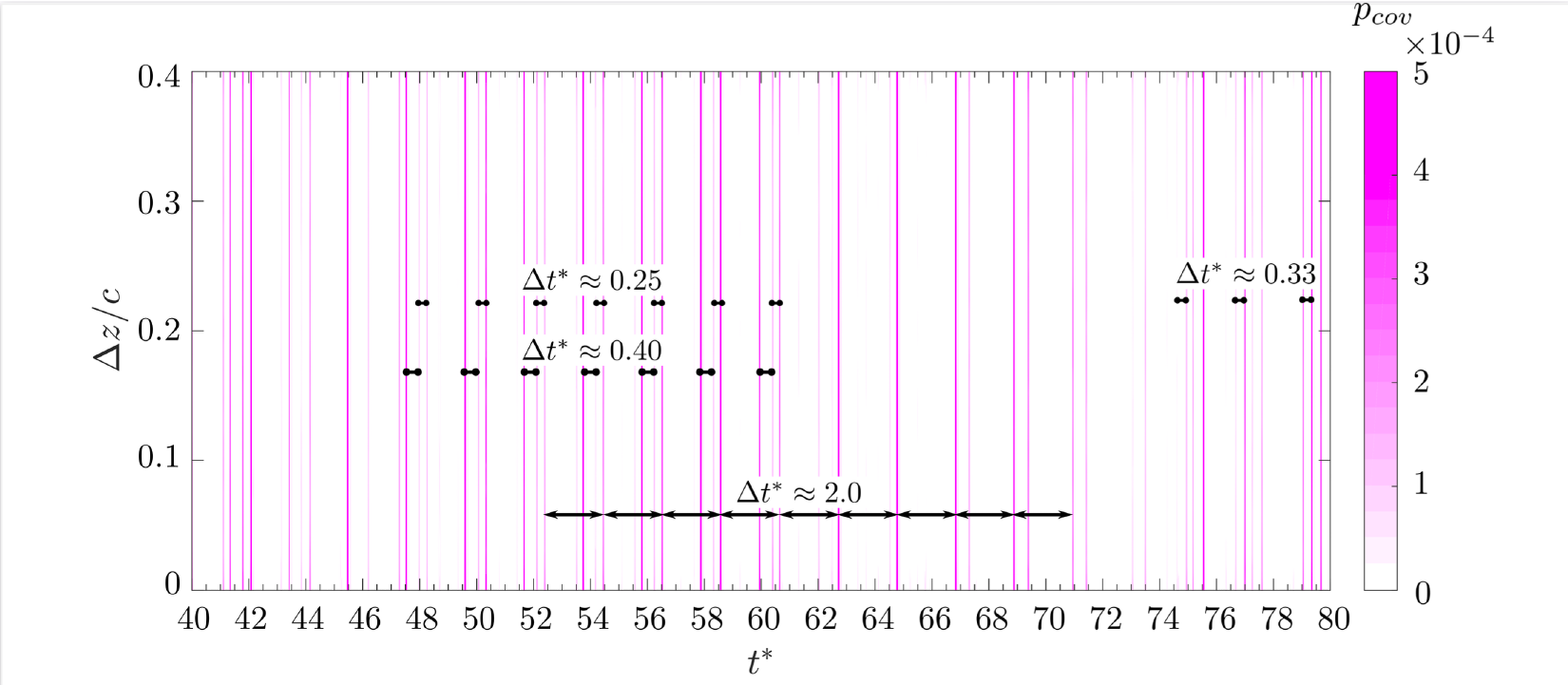}
\caption{Spatial covariance of pressure fluctuations $p'$ computed near the airfoil trailing edge, on the suction side. Different intervals of $\Delta t^*$ are depicted in the figure highlighting the intermittent behavior of 2D vortex shedding.}
\label{f:covariance} 
\end{figure}

The study of vortex dynamics outlined in Section \ref{s:vortex} displays  distinct patterns of flow structures shed from the LSB. Here, we assess the intermittent behavior of the coherent structures as they reach the TE.
Figure \ref{f:covariance} shows the spanwise covariance of the pressure fluctuations calculated at $x/c=0.98$ and a distance $\Delta n/c=0.02$ from the wall, i.e., near the trailing edge, for different time instants as
\begin{equation}
p_{cov}(\Delta z,t)=\langle p'(x,y,z,t)\, p'(x,y,z+\Delta z,t)  \rangle \mbox{ .}
\end{equation}
Thicker magenta lines indicate stronger instantaneous covariance values along the span, representing the shedding of quasi-2D vortices that are responsible for the strong acoustic emission shown in the top-right plot of Fig. \ref{f:acfields}. Regions of white color represent instants when the flow is either turbulent or laminar but steady, without any correlated structure.

The intermittent shedding of the coherent structures can be observed by the different time intervals marked in the figure. Several structures reach the TE with time intervals of $\Delta t^* \approx 2.0$, which is equivalent to a Strouhal number $St = fc/U_\infty \approx 0.5$, where $f$ is the shedding frequency. Other vortices are shed at faster time scales such as $\Delta t^* \approx 0.25$, $0.33$, and $0.40$, which result in nondimensional frequencies of $St \approx 4.0$, $3.0$, and $2.5$, respectively. It is noteworthy to mention that these shedding frequencies could also be similarly extracted from Fig. \ref{f:Cfanalysis} if the frequency occurrence of the different stripes of $\langle C_f \rangle$ near the TE are to be quantified.

The different types of events depicted in Fig. \ref{f:covariance} demonstrate a switch between 3D and 2D-like flow structures. The former are due to instants when the flow is turbulent, while the latter can be related to spanwise-coherent laminar vortices. To investigate the role of these different events on the present unsteady flow, we follow Ref. \cite{larue1974} and employ an intermittency function $I(t^*)= H(\langle p_{cov}(t^*) \rangle - p_{th})$ to distinguish the 2D from the 3D flow events. Here, $H$ is the Heaviside function, $\langle p_{cov}(t^*) \rangle$ is the instantaneous spanwise-averaged pressure covariance, and $p_{th}$ is a threshold value based on the maximum level of Fig. \ref{f:covariance}, chosen as $p_{th}=5\times10^{-4}$ to separate the events. This threshold is set after investigating how the 2D and 3D events are split with respect to function $I$. It is noticed that the current value of $p_{th}$ provides a clear separation of events for the position considered in the covariance calculation. Through the intermittency function $I$, the intermittency factor $\gamma_{_{I}}(T)$ can be calculated as:
\begin{equation}
\gamma_{_{I}}(T) = \frac{1}{T}
\int_{0}^{T} I(t^*) \mathrm{d} t^*
\mbox{ ,}
\end{equation}
where $T$ is the total period of the simulation.

The individual contributions of the 2D and 3D events to the mean flow quantities $\overline{q}$ are computed as 
\begin{equation}
\label{eqn:mean_turb}
\overline{q}_{2D} = \frac{1}{\gamma_{_{I}} T} \int_{0}^{T} q(t^*)I(t^*) \mathrm{d} t^* \mbox{ ,}
\end{equation}
and
\begin{equation}
\label{eqn:mean_nturb}
\overline{q}_{3D} = \frac{1}{(1-\gamma_{_{I}}) T} \int_{0}^{T} q(t^*)[1-I(t^*)] \mathrm{d} t^* \mbox{ .}
\end{equation}
Similar calculations can be performed to obtain the contributions of the 2D and 3D events to the fluctuation quantities as $q'$
\begin{equation}
\label{eqn:fluc_turb}
\overline{q'^2}_{2D} = \frac{1}{\gamma_{_{I}} T} \int_{0}^{T} [q(t^*)- \overline{q}_{2D}]^2 I(t^*) \mathrm{d} t^* \mbox{ ,}
\end{equation}
and
\begin{equation}
\label{eqn:fluc_nturb}
\overline{q'^2}_{3D} = \frac{1}{(1-\gamma_{_{I}}) T} \int_{0}^{T} [q(t^*)- \overline{q}_{3D}]^2 [1-I(t^*)] \mathrm{d} t^* \mbox{ .}
\end{equation}
In the above equations, $q$ represents any property of the flow.

\begin{figure}[h!]
\centering
   	\begin{subfigure}{0.94\textwidth}
 	\centering
       	\includegraphics[trim={0.2cm 0 0.3cm 0},clip,width=1\linewidth]{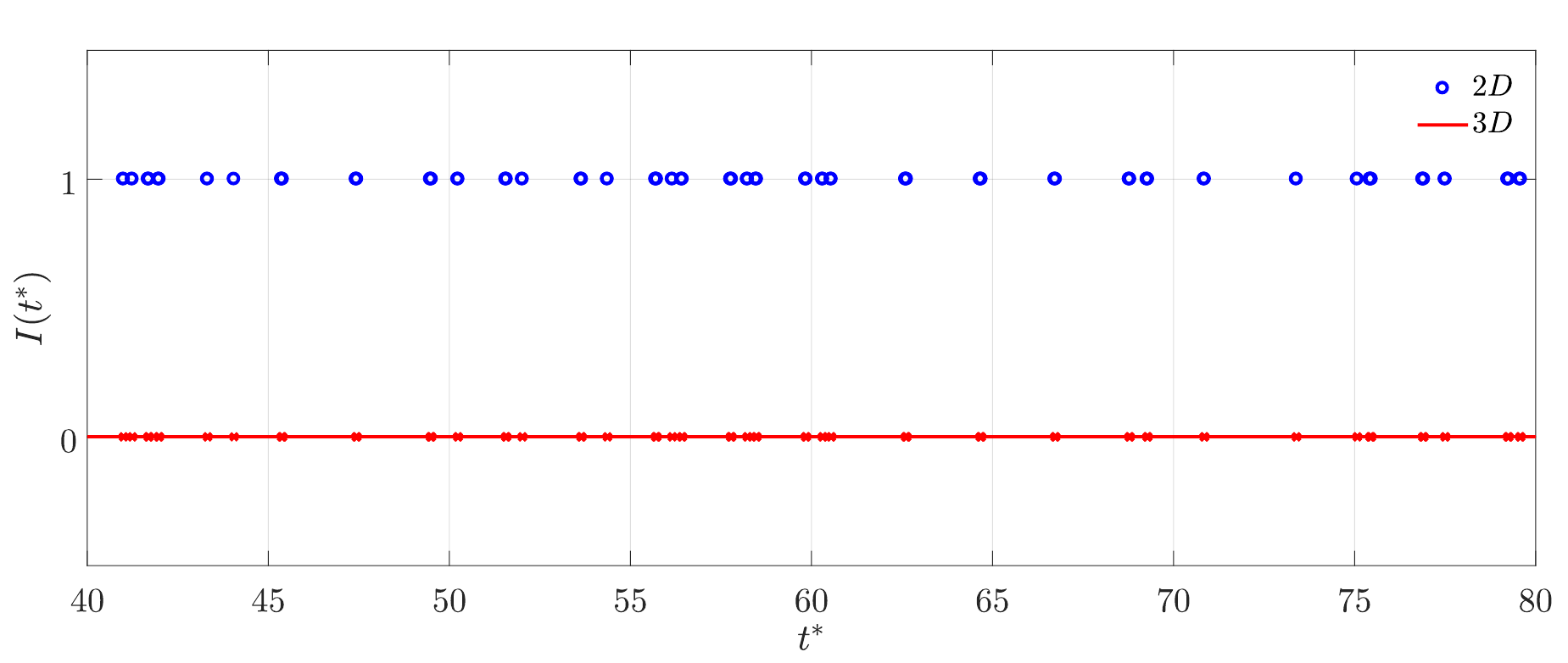}
 		\caption{}
 	      \label{f:funcI}
 	\end{subfigure}  
   	\begin{subfigure}{0.49\textwidth}
 	\centering
       	\includegraphics[trim={0 0 0cm 0},clip,width=1\linewidth]{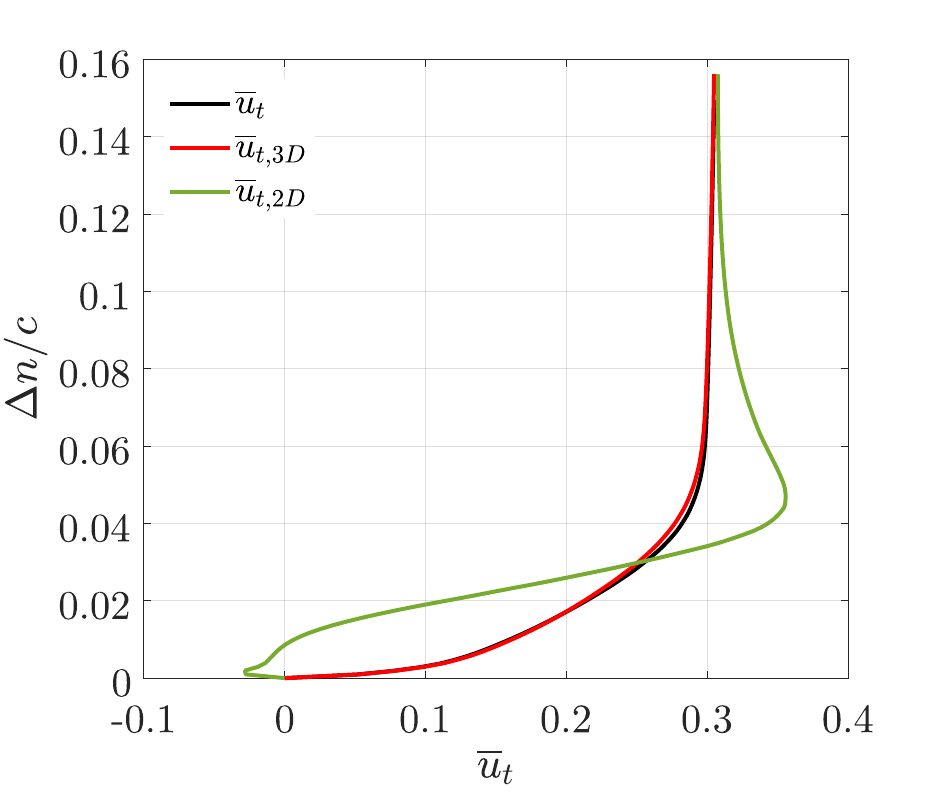}
 		\caption{}
 	      \label{f:utavg}
 	\end{subfigure}    
   	\begin{subfigure}{0.49\textwidth}
 	\centering
       	\includegraphics[trim={0 0 0cm 0},clip,width=1\linewidth]{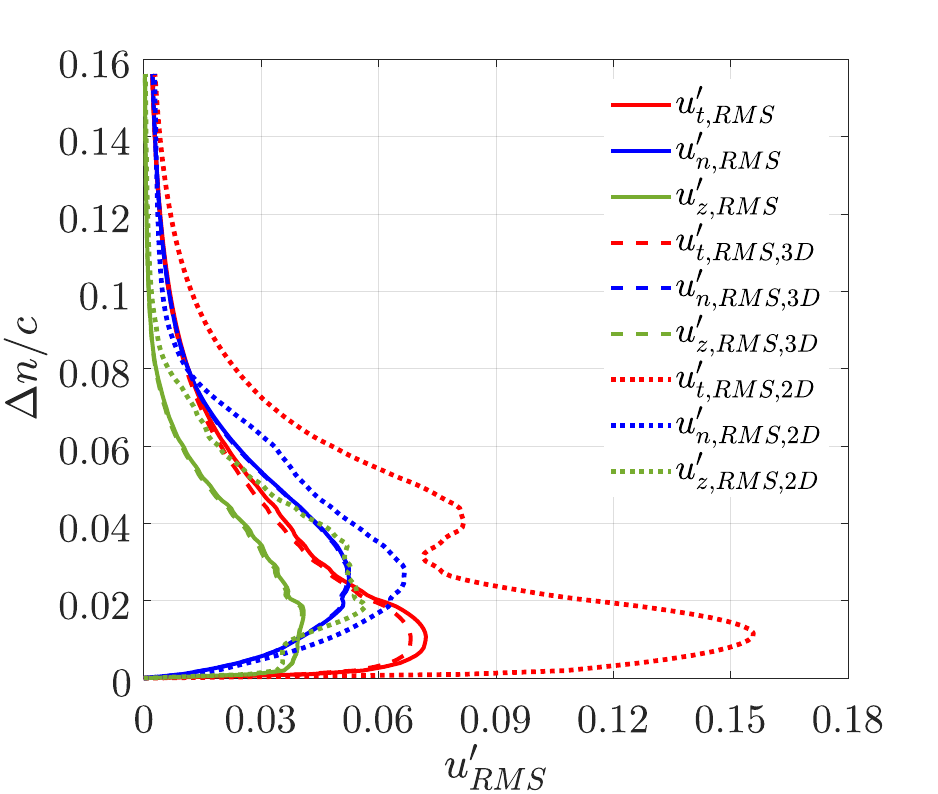}
 		\caption{}
 	      \label{f:uRMS}
 	\end{subfigure}      
\caption{Intermittency analysis. (a) Intermittency function $I(t^*)$ which divides the 2D and 3D events; (b) Time- and spanwise-averaged tangential velocity, $\overline{u}_t$, displayed for the 3D and 2D flow contributions; (c) RMS values of the tangential, wall-normal and spanwise velocities along the normal direction considered.}
\label{f:interm} 
\end{figure}

The $I$ function is displayed in Fig. \ref{f:funcI} where the blue markers depict the instants when the spanwise-averaged spatial covariance is above the threshold. In contrast, the red line represents the 3D events and the respective red markers represent the same instants of the 2D events. The intermittency analysis is performed at a normal line from point P4 (see Fig. \ref{fig:points}), a region where intermittency is expected after the BL is reattached to the airfoil. The split between 2D and 3D events are first depicted through the mean tangential velocity profile with respect to the normal direction in Fig. \ref{f:utavg}. In the majority of time, the flow can be considered 3D since the line corresponding to the mean profile of the full flow field is practically superposed with the $\overline{u}_{t,3D}$ line. Nevertheless, the green line shows the effect of the 2D structures on the mean velocity and close to the airfoil ($\Delta n/c < 0.01$) flow separation occurs ($\overline{u}_t < 0$) and it is solely captured by these 2D events. This observation confirms the analysis of $\langle C_f \rangle$ from Fig. \ref{f:Cfanalysis}, where spanwise-coherent structures of negative z-vorticity are shown to generate positive z-vorticity near the wall.
Above $\Delta n/c > 0.03$, the mean velocity is higher than that of the freestream ($\overline{u}_t > 0.3$). Here, velocity results are presented normalized by the speed of sound. 

To investigate the fluctuation of the velocity components induced by the 2D structures over the airfoil, the RMS values of the tangential, wall-normal, and spanwise velocities are depicted in Fig. \ref{f:uRMS}. The tangential component dominates over the wall-normal and spanwise ones due to being closer to the freestream velocity direction. The $u'_{t,RMS,2D}$ line shows two peaks that are possibly due to the vortex coherence effect taking place within $0 < \Delta n/c < 0.05$ from the foil surface. Similar to the mean profiles, the velocity fluctuations can be considered 3D most of the time since for all velocity components, the profiles corresponding to the full flow field almost overlap with the 3D ones. The surprising result here is the contribution of the spanwise component. Despite the 2D vortex and the spanwise periodic boundary conditions, results show that the z-velocity component is not negligible and yet the $u'_{z,RMS,2D}$ line closely matches the normal component considering the full flow field ($u'_{n,RMS}$). 

\FloatBarrier

\begin{figure}[h!]
\centering
        \includegraphics[width=0.9\textwidth, trim={0 0 0 0 0}, clip]{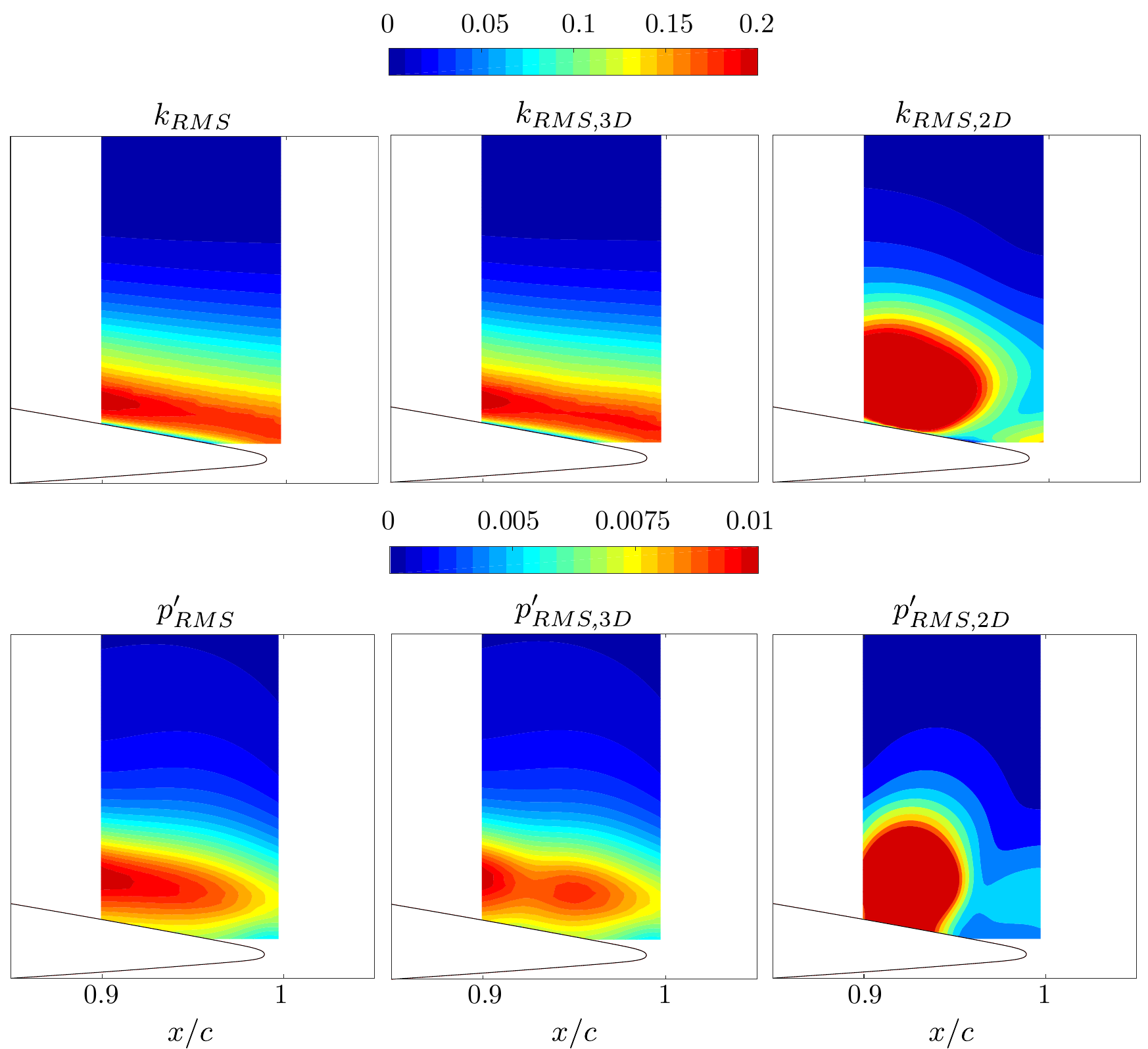}
\caption{RMS flow fields of turbulent kinetic energy and pressure fluctuation for the full flow field (left column), and the separate 3D (center) and 2D (right) flow fields.}
\label{f:upRMSfield} 
\end{figure}

To delve into the RMS results from Fig. \ref{f:uRMS} and analyze how it varies spatially in the region of intermittency, the flow fields corresponding to the turbulent kinetic energy and pressure fluctuations are shown in Fig. \ref{f:upRMSfield}. The takeaways here are the similarity between full and 3D flow fields (first and second columns) and the circular shape displayed in the third column for the 2D flow field, resembling a vortex. One can also see that the 2D fluctuations are considerably stronger than the 3D ones, as previously outlined in Fig. \ref{f:uRMS}. For instance, the 2D pressure fluctuations near the trailing edge are more intense than those induced by the 3D flow fields. This confirms the importance of the spanwise-coherent events to the trailing edge noise generation.

\FloatBarrier

\subsection{Spectral analysis}

The flow intermittency analysis of Section \ref{s:interm} described the contribution of 2D and 3D events in the velocity and pressure fluctuations on the airfoil TE. The frequency and amplitude modulations of these fluctuations play an important role in airfoil transitional flows \cite{Desquesnes2007, probsting2014, Ricciardi2019_tones}. As observed from the instantaneous spanwise-averaged friction coefficient map of Fig. \ref{f:Cfanalysis}, the bubble located on the airfoil suction side modulates the shedding of Kelvin-Helmholtz instabilities that are advected toward the trailing edge. According to \cite{ricciardiJFM}, this modulation is shown to impact the generation of spanwise-coherent or uncorrelated turbulent structures via constructive/destructive interference of the vortex shedding frequencies.

\begin{figure}[h!]
    \centering
    \begin{subfigure}[b]{0.78\textwidth}
        \includegraphics[trim={1.3cm 0 2cm 0.5cm},clip,width=\linewidth]{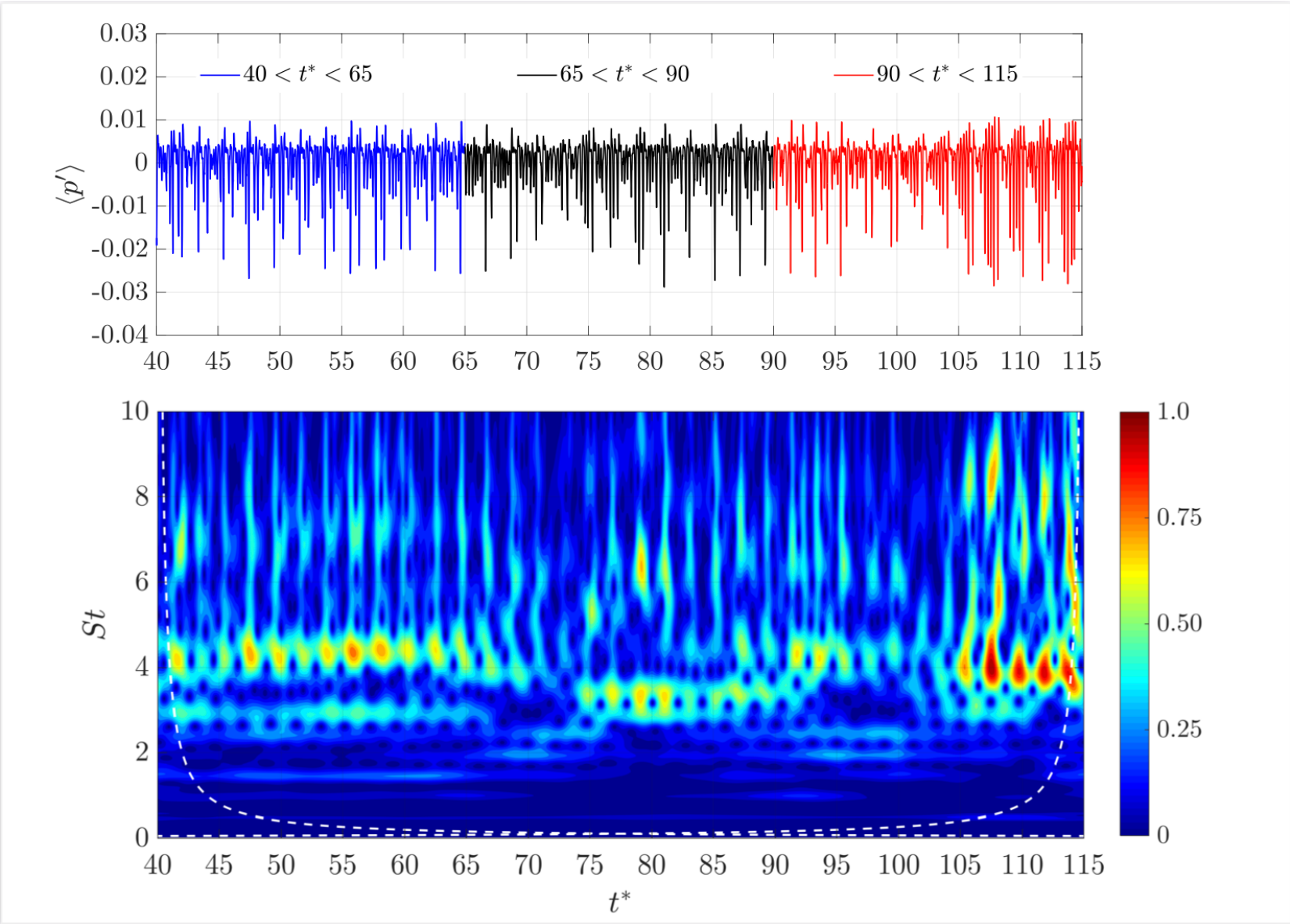}
        \caption{LBM}
        \label{wavelet1}
    \end{subfigure}
    \begin{subfigure}[b]{0.78\textwidth}
        \includegraphics[trim={1.3cm 0 2cm 0.5cm},clip,width=\linewidth]{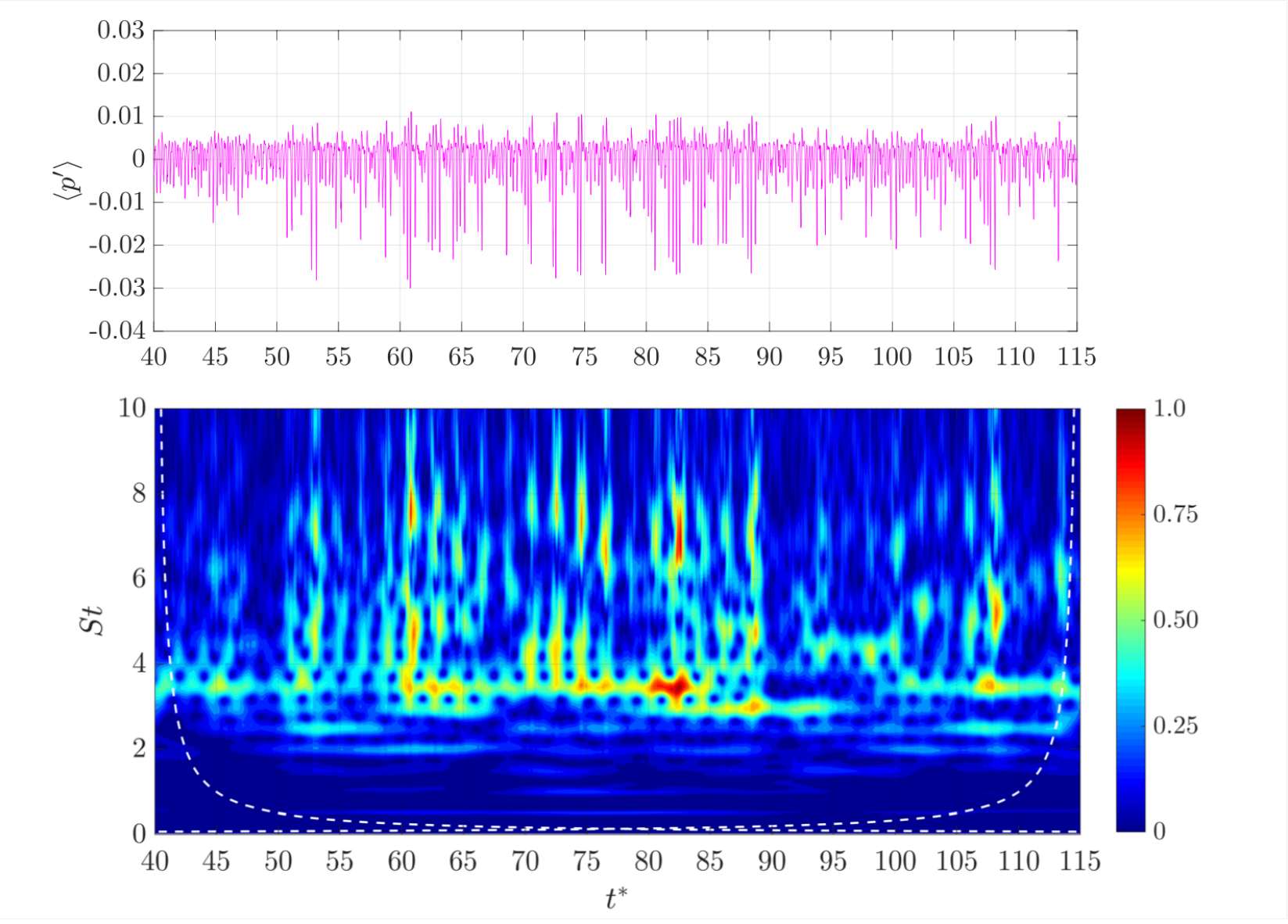}
        \caption{NS}
        \label{wavelet2}
    \end{subfigure}
    \caption{Temporal signals of spanwise-averaged pressure fluctuations, computed at $x/c = 0.98$ and $\Delta n/c = 0.02$, and time-frequency analyses of pressure fluctuations using a Morlet wavelet.}
    \label{wavelet}
\end{figure}

A spectral analysis of the present flow provides insights of the flow dynamics and noise generation mechanisms identifying the instantaneous frequencies excited near the TE. 
Therefore, hydrodynamic pressure fluctuations are extracted on the airfoil suction side at $x/c = 0.98$, at a distance $\Delta n/c = 0.02$ from the airfoil suction side. The temporal signal computed by the baseline mesh of the LBM is shown in the top plot of Fig. \ref{wavelet1}, displaying strong negative pressure peaks related to the intermittent passage of spanwise-correlated vortical structures. The signal also shows smaller amplitude oscillations related to turbulent packets and it is divided into bins of 25$t^*$ each, marked by different colors according to the legend. Such bin differentiation is utilized after performing a continuous wavelet transform (CWT) \cite{farge1992}.


Here, a time-frequency analysis is performed by using a Morlet wavelet to characterize the amplitude and frequency modulation effects that occur in the present flow. This wavelet is composed of a monochromatic complex exponential multiplied by a Gaussian envelope, and it provides a trade-off in terms of temporal and frequency resolutions. Spectrograms are presented in Fig. \ref{wavelet} for the LBM (Fig. \ref{wavelet1}) and NS (Fig. \ref{wavelet2}) simulations, showing the correlation of the wavelet at different instants and frequencies with the respective pressure fluctuation signal. A higher wavelet magnitude indicates stronger fluctuations at a particular frequency and time. Since the wavelet magnitude depends on the standard deviation of the signal and its reference frequency \cite{farge1992}, and to focus on the intermittency aspects of the signals computed for the LBM and NS approaches, the magnitude here is normalized by the maximum value achieved by each methodology.

The LBM wavelet exhibits three distinct patterns according to the dominant shedding frequency. The first one (blue color line in Fig. \ref{wavelet1}) has a time range $40<t^*<65$ and a dominant frequency of $St \approx 4.3$. The second pattern (black color) has a dominant peak at $St \approx 3.3$ and ranges from $65<t^*<90$. Lastly, an intermediate Strouhal number of $St \approx 3.8$ appears in the last $25t^*$ depicting stronger pressure fluctuations compared to the previous patterns. In contrast, the NS spectrogram has a more uniform pattern across all $75t^*$ with the main tone at $St \approx 3.3$. Overall, the pressure fluctuation signal from the LBM appears to be more intermittent compared to the NS due to the main tone switch during the simulation.



\FloatBarrier

\begin{figure}[h!]
    \centering
    \begin{subfigure}[b]{0.49\textwidth}
        \includegraphics[trim={0 0 1.4cm 0},clip,width=\linewidth]{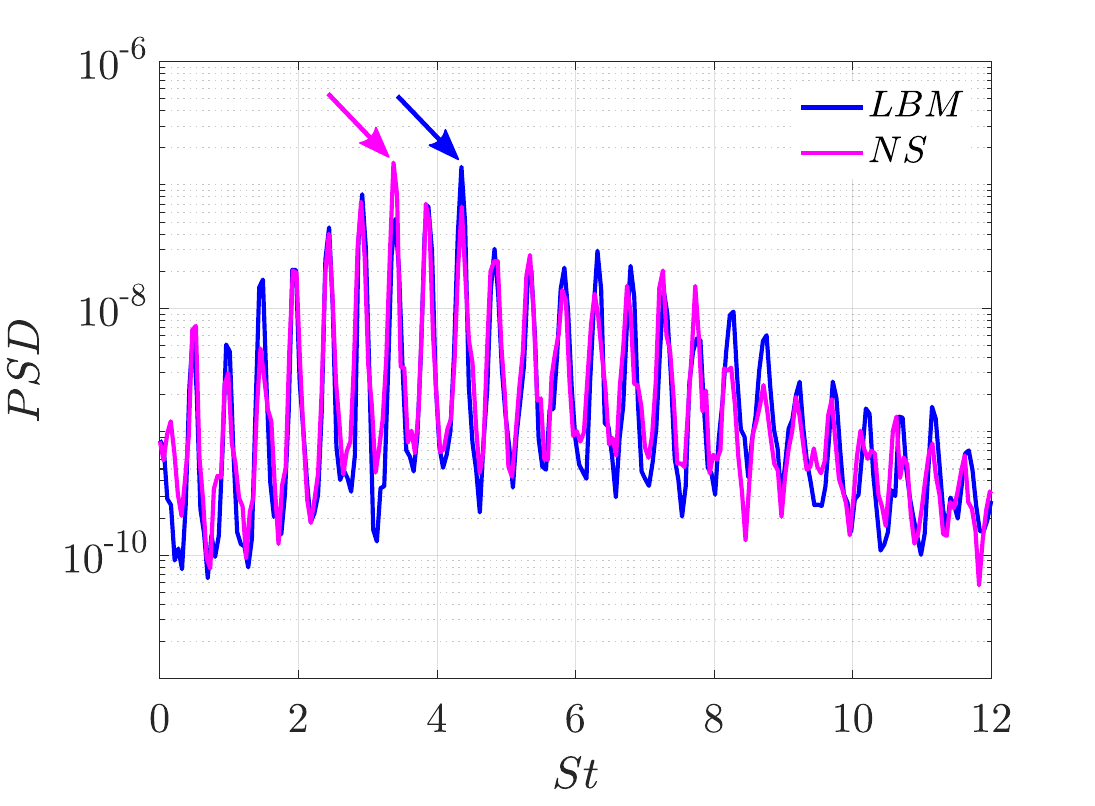}
        \caption{}
        \label{psd1}
    \end{subfigure}
    \hfill
    \begin{subfigure}[b]{0.49\textwidth}
        \includegraphics[trim={0 0 1.4cm 0},clip,width=\linewidth]{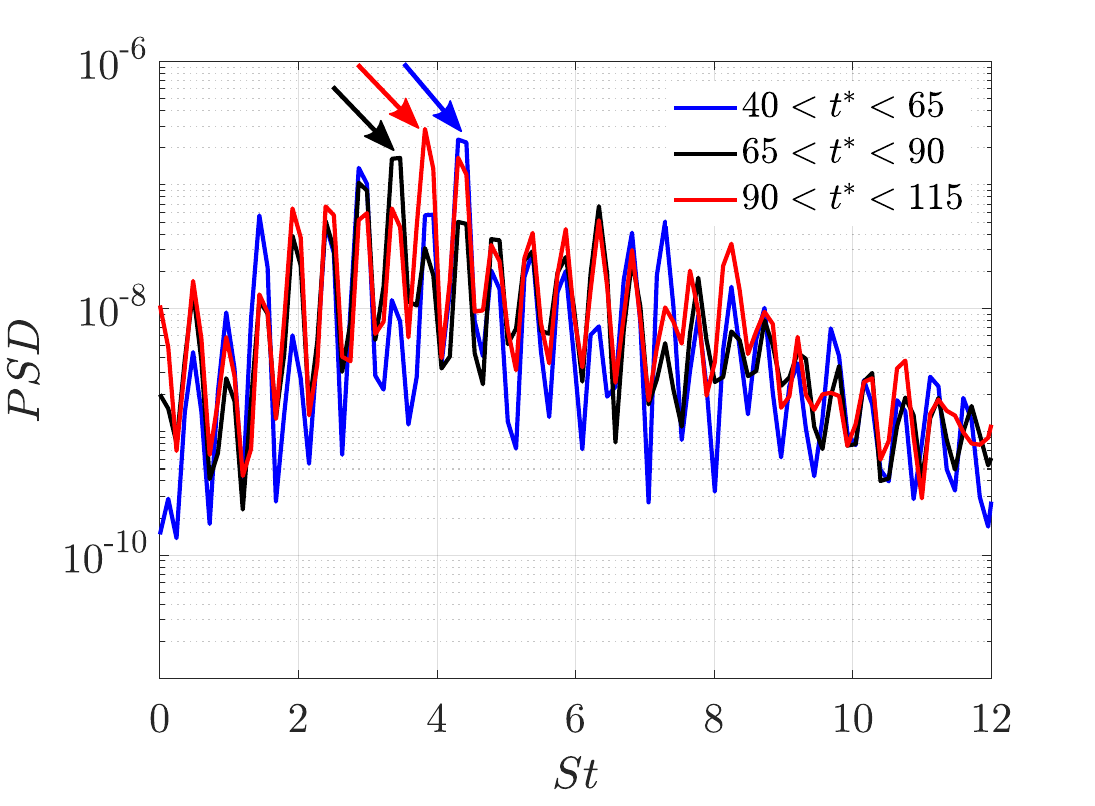}
        \caption{}
        \label{psd2}
    \end{subfigure}
    \caption{PSD analysis for (a) the full time signal of the pressure fluctuation between LBM and NS at $x/c = 0.98$ and $\Delta n/c = 0.02$, and (b) across the three LBM wavelet patterns shown in Fig. \ref{wavelet1}.}
    \label{psd}
\end{figure}

The wavelet analyses showed that the main tones of the present flow may switch frequencies in time due to the intermittent vortex shedding from the separation bubble.
To quantify and compare the main and secondary tones of the pressure fluctuation signals showed in Fig. \ref{wavelet} between the LBM and NS, power spectral densities (PSDs) are computed for the entire spanwise-averaged signals. A fast Fourier transform (FFT) is employed to the full time signal of the LBM and NS calculations. Both signals are divided in $7$ bins with $50\%$ of overlap and the PSDs are shown in Fig. \ref{psd1}. The LBM and NS datasets have $8712$ and $5000$ snapshots, respectively. To force signal periodicity and reduce spectral leakage, a Hanning window is used in both pressure data. As can be observed, there is a remarkable similarity in the PSDs of both methodologies with the only major difference being the main tone frequency of $St \approx 4.3$ for the LBM (blue arrow) and $St \approx 3.3$ for the NS (magenta arrow). 

In both LBM and NS, all tonal peaks are integer multiples of the lowest-frequency tone, $St \approx 0.48$, where the strength of the secondary tones are also similar for the different approaches. These frequencies are related to the passage of low-pressure disturbances, either turbulent or coherent, near the TE. A PSD analysis is also performed to each of the three different temporal patterns of the LBM signal. Here, the signals are divided into 5 bins with 50\% overlap. As depicted in Fig. \ref{psd2}, three different main tones are captured depending on the sample, and the same main tone of the NS approach is observed for the intermediate sample with $65<t^*<90$. In addition to the main tones, as already observed in Fig. \ref{wavelet1}, the PSD shows that the magnitudes of the secondary tones vary significantly with respect to the time samples. The spectral analysis shows that the present intermittent flow depicts strong amplitude and frequency modulations that cause the main tone to switch frequencies in time.   


\FloatBarrier

\section{Conclusions}

The study of an intermittent transitional airfoil flow with a laminar separation bubble is performed using the LBM. Results of a flow over a NACA0012 airfoil with Reynolds number $Re=5 \times 10^4$, freestream Mach number of $M_\infty=0.3$, and angle of attack $\alpha=3^{\circ}$ are compared and validated against a wall-resolved LES of the NS equations. Good agreement is observed between the different approaches in terms of mean and RMS flow quantities. The separation and reattachment positions of the LSB are found to be the same for the LBM and NS solutions. Power spectral densities of pressure fluctuations at the trailing edge also show remarkable similarities in terms of tone frequencies and magnitudes.

Intermittent vortex shedding from the LSB impacts the unsteady aerodynamics and trailing-edge noise generation. Visualization of instantaneous spanwise vorticity shows different patterns of structures shed from the LSB, being single vortices that keep coherence until the trailing edge, or that pair with other vortices also keeping coherence, besides their counterparts that break into finer turbulent scales. These vortical structures impress different patterns of instantaneous spanwise-averaged skin friction over the airfoil. Results also demonstrate that the LBM is able to resolve small-scale vorticity dynamics near the wall. Temporal analysis of pressure spanwise coherence shows that the trailing edge dynamics is characterized by intermittent 3D and 2D events. The latter ones are related to stronger RMS fluctuations near the trailing edge which, in turn, generate stronger pressure pulses due to acoustic scattering. Lastly, results of wavelet analyses and PSDs applied at a point near the airfoil TE show that the main tone frequency from the LBM simulation switches with respect to time. The nondimensional frequencies range from $St \approx 3.3$ to $St \approx 4.3$, being a multiple of the lowest frequency tone at $St \approx 0.48$. The magnitudes of the main and secondary tones are also impacted by the intermittency of the vortex shedding from the LSB. This research provides insights on the flow dynamics around blades and propellers used in MAVs and eVTOLs, which have been shown to develop LSBs and generate trailing-edge noise. 

\section*{Acknowledgments}

The authors thank CENAPAD-SP (Cluster Lovelace) for providing the computational resources used in this study (project 551). We also acknowledge Fundação de Amparo à Pesquisa do Estado de São Paulo (FAPESP) for supporting the present work under research grants No. 2013/08293-7, 2021/06448-0, and 2023/08768-7. The second author is supported by a scholarship from Coordenação de Aperfeiçoamento de Pessoal de Nível Superior (CAPES) under grant 88887.825328/2023-00.


\bibliographystyle{elsarticle-num} 
\bibliography{cas-refs}

\end{document}